\documentclass[namedreferences]{solarphysics}
\usepackage[hyperref,optionalrh,showbiblabels]{style} 
\usepackage{graphicx}                    
\usepackage{color}                       
\usepackage{breakurl}                    

\usepackage{siunitx} 
\usepackage{placeins}
\usepackage{float}
\usepackage{caption} 
\usepackage{booktabs}
\usepackage[T1]{fontenc}

\chardef\us=`\_

\begin{document}

\begin{article}

\begin{opening}

\title{Differential Emission Measure Plasma Diagnostics of a Long-Lived Coronal Hole}
		\author[addressref={aff1}]{\inits{J.S}\fnm{Jonas}~\lnm{Saqri}\orcid{0000-0003-2170-1140}}
		\author[addressref={aff1},corref,email={astrid.veronig@uni-graz.at}]{\inits{A.V}\fnm{Astrid~M.}~\lnm{Veronig}\orcid{0000-0003-2073-002X}}
		\author[addressref=aff1]{ \inits{S.H}\fnm{Stephan~G.}~\lnm{Heinemann}\orcid{0000-0002-2655-2108}}
		\author[addressref=aff1]{\inits{S.H}\fnm{Stefan~J.}~\lnm{Hofmeister}\orcid{0000-0001-7662-1960}}
		\author[addressref=aff1]{\inits{M.T}\fnm{Manuela}~\lnm{Temmer}\orcid{0000-0003-4867-7558}}
		\author[addressref=aff1]{\inits{K.D}\fnm{Karin}~\lnm{Dissauer}\orcid{0000-0001-5661-9759}}
		\author[addressref=aff2]{\inits{Y.S}\fnm{Yang}~\lnm{Su}\orcid{0000-0002-4241-9921}}
%
\runningauthor{J. Saqri {\em et al.}}
\runningtitle{DEM Plasma Diagnostics of a Long-Lived Coronal Hole}
\address[id=aff1]{Institute of Physics \& Kanzelh\"ohe Observatory, University of Graz, Austria}
\address[id=aff2]{Purple Mountain Observatory, Chinese Academy of Sciences (CAS), Nanjing, China}
\begin{abstract}
	We use \textit{Solar Dynamics Observatory} (\textit{SDO})/\textit{Atmospheric Imaging Assembly} (\textit{AIA}) data to reconstruct the plasma properties from differential emission measure (DEM) analysis for a previously studied long-lived, low-latitude coronal hole (CH) over its lifetime of ten solar rotations. We initially obtain a non-isothermal DEM distribution with a dominant component centered around 0.9\,MK and a secondary smaller component at 1.5\,--\,2.0\,MK. We find that deconvolving the data with the instrument point spread function (PSF) to account for long-range scattered light reduces the secondary hot component. Using the 2012 Venus transit and a 2013 lunar eclipse to test the efficiency of this deconvolution, significant amounts of residual stray light are found for the occulted areas. Accounting for this stray light in the error budget of the different \textit{AIA} filters further reduces the secondary hot emission, yielding CH DEM distributions that are close to isothermal with the main contribution centered around 0.9\,MK. Based on these DEMs, we analyze the evolution of the emission measure (EM), density, and averaged temperature during the CH's lifetime. We find that once the CH is clearly observed in EUV images, the bulk of the CH plasma reveals a quite constant state, \textit{i.e.} temperature and density reveal no major changes, whereas the total CH area and the photospheric magnetic fine structure inside the CH show a distinct evolutionary pattern. These findings suggest that CH plasma properties are mostly ``set'' at the CH formation or/and that all CHs have similar plasma properties.
		\end{abstract}
%
\end{opening}

	\section{Introduction}
	\label{S-Introduction} 
	Coronal holes are regions of reduced plasma density and temperature in the solar corona.
	Thus they appear darker than the surrounding quiet Sun in extreme-ultraviolet (EUV) and X-ray images.
	Magnetically, CHs are characterized by the prevailing dominant polarity nature of the underlying magnetic field where magnetic-field lines of the dominant polarity are ``open" into interplanetary space \citep{wilcox1968,altschuler1972,hundhausen1972coronal}. Along these open magnetic-field lines, plasma is accelerated, escaping the Sun's gravitational field forming interplanetary high-speed solar wind streams. The solar wind parameters speed and proton temperature measured at 1\,AU are correlated with the CH area \citep{Krieger1973,vrsnak2007,Rotter2012}.
	\par
	These open flux tubes are believed to be rooted in concentrations of magnetic flux, known as magnetic elements \citep{hassler1999,tu2005}.
	Based on a study of 288 low-latitude CHs, \cite{hofmeister2017} found that these magnetic elements are also the main origin of the unbalanced magnetic flux of CHs. The total area covered by long-lived magnetic elements  with lifetimes \textgreater~four days is strongly correlated with the unbalanced magnetic flux of CHs (cc=0.99). These magnetic elements account for $\approx$ 68\,\% of the unbalanced ``open" magnetic flux of CHs at typical coverages of about 3\,\% of the total CH area \citep{hofmeister2019,2019statCH}.
	\par
	\cite{heinemann2018A,heinemann2018B} investigated a long-lived low-latitude CH observed by the \textit{Solar Dynamics Observatory} (\textit{SDO}) and \textit{Solar Terrestrial Relations Observatory} (\textit{STEREO}) spacecraft over ten solar rotations from February 2012 to October 2012 (see Figure. \ref{F-CH193A}). Based on the CH area, they identified a distinct three-phase evolution characterized by a growing, maximum, and decaying phase. For this CH, a strong correlation between CH area and the mean magnetic-field strength inside the CH was found. The signed magnetic-field strength increased over the course of the growing phase and reached its maximum at the time of maximal CH area followed by a steep drop in field strength marking the beginning of the CH's decaying phase.
	During the maximum phase, a larger fraction of the overall magnetic flux is concentrated in magnetic elements than in the growing and decaying phases, and the number of magnetic elements per area for elements with field strengths \textgreater 50\,G increases nearly 100\,\% compared to either the beginning or end of the CH's life. This maximum phase further exhibits the strongest correlation between CH area and peak speed of the associated high-speed solar wind stream.
	\par
	As the magnetic-field configuration is assumed to be responsible both for the coronal heating (\textit{e.g.} \citealp{abr2006,klimchuck2006}) and for the acceleration of the solar-wind plasma \citep{wang2010}, the clear trend in the magnetic structure and the distinct correlations with the fast solar-wind characteristics of this CH identified by \cite{heinemann2018A,heinemann2018B} make it an ideal candidate for probing the effect of changing magnetic fine structure on CH plasma.
	The low-latitude position of the CH under study minimizes projection effects, thus making it also a suitable candidate for differential emission measure (DEM) analysis. This allows us to deduce the temperature-dependent emission of the optically thin coronal plasma by studying multi-wavelength EUV measurements in combination with elemental abundances and the wavelength dependent instrument response.
	\par 
	In this article we apply DEM diagnostics using the inversion code developed by \cite{HK2012} using \textit{SDO}/\textit{Atmospheric Imaging Assembly} (\textit{AIA}) data to reconstruct the DEM distribution of a CH and infer the density and temperature evolution over its lifetime.
\section{Data and Methods}
\label{S-DataMethods}  
	\subsection{Data}
	We made use of six of the seven EUV channels (94, 131, 171, 193, 211 and 335\,\AA) of the \textit{AIA} instrument \citep{Lemen2012} onboard the \textit{SDO} \citep{pesnell2012} spacecraft, sensitive over a temperature range from $10^{5}$\,K to $10^{7}$\,K. The 304\,\AA \ filter was omitted because it images optically thick plasma making it unsuitable for DEM analysis, which relies on optically thin emission. With the exception of the He \textsc{II} imaging 304\,\AA \ channel, \textit{AIA's} EUV filters primarily observe different Fe ions (with ionisation states from Fe \textsc{VIII} to Fe \textsc{XXIV}).
	\par
\begin{figure}[t]   
	\centering
	\includegraphics[width=1.0\textwidth,clip=]{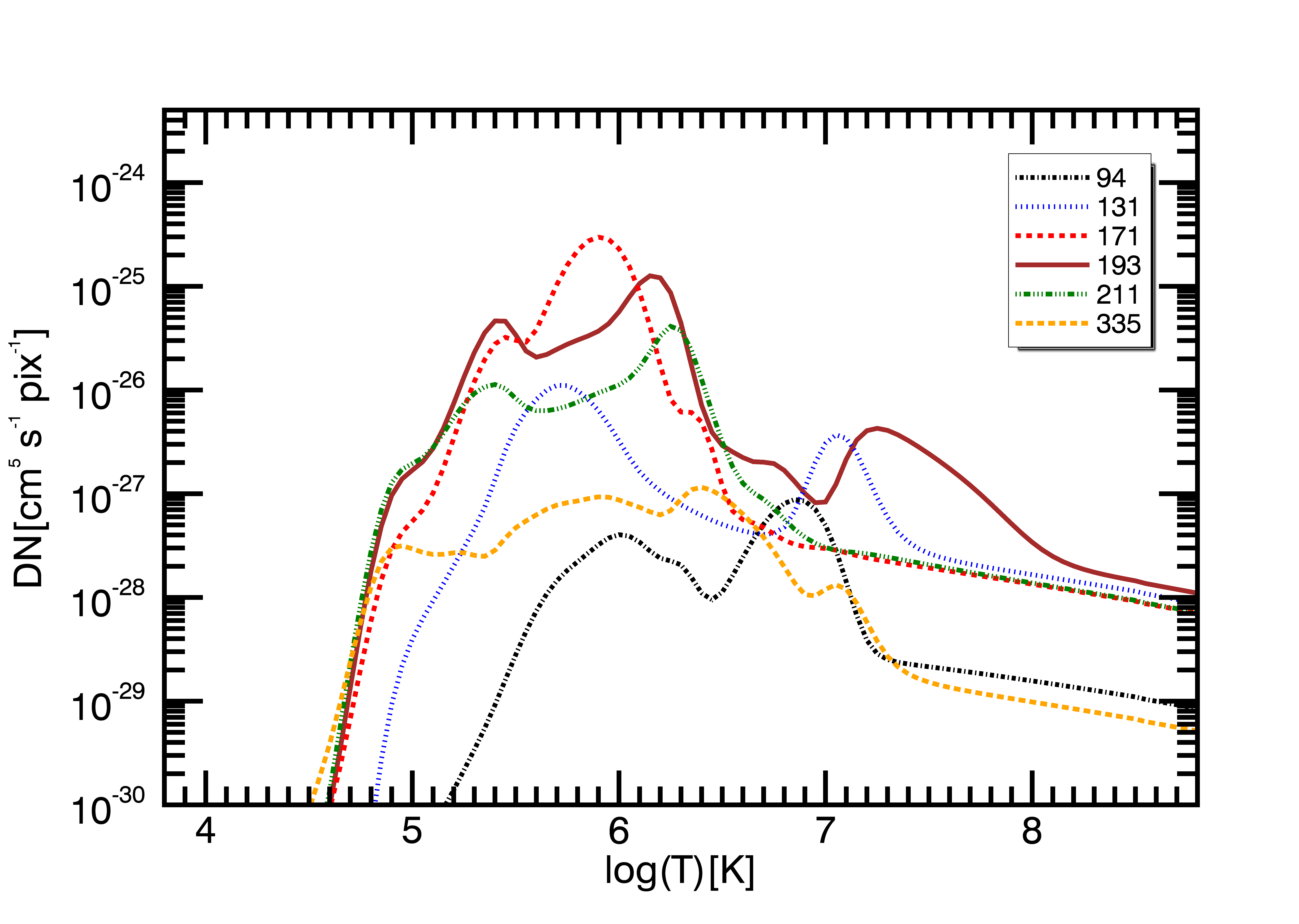}
	\caption{Temperature response functions for the utilized \textit{AIA} EUV filters generated using CHIANTI 9 and the effective areas provided by $\sf{aia\_get\_response.pro}$.}
	\label{F-AIAFilterResponse}
	\end{figure}

	The \textit{AIA} temperature response functions of the filters utilized are given in Figure \ref{F-AIAFilterResponse}.
	The AIA pixel scale is 0.6 arcsec \citep{Lemen2012}. To enhance the signal-to-noise ratio, the data were binned by 8$\times$8 pixels while conserving flux, resulting in an effective spatial resolution of 4.8\,arcsec per pixel. The \textit{AIA} data were deconvolved with the instrument point spread function (PSF) in order to reduce the instrument stray light and its effect on the reconstructed DEMs (Section \ref{sect:Deconv}). The data were further processed using the standard SSWIDL data reduction routine \textsf{aia\_prep.pro}.
	\par
	We then perform the DEM analysis at a cadence of one hour during the $\pm$ six-hour time window where the CH is nearest to the disk center to reduce LOS-integration effects of the emisson of close-by structures. We applied the same CH boundaries as in \cite{heinemann2018A,heinemann2018B} which were generated using an intensity-based threshold method applied to the \textit{AIA} 193\,\AA \ images \citep{Rotter2012,hofmeister2017}.
	\subsection{Differential Emission Measure Analysis}
	\label{subsect:DEMAnalysis}
	For optically thin emission from plasma in thermodynamic equilibrium the temperature distribution of the contributing plasma in the LOS is described by the differential emission measure defined as
	\begin{equation} 
	\label{Eq-DEM-def}
		\mathrm{DEM(}T\mathrm{)} = n(T)^{2}\frac{\mathrm{d}h}{\mathrm{d}T},
	\end{equation}
	where $n$ is the number density dependent on the temperature $T$ along the LOS. The measured intensity $I_{\lambda}(T)$ for a given \textit{AIA} filter is then related to the DEM via 
	\begin{equation}
	\label{Eq-IDEM-def}
		I_{\lambda} =\int_{T} K_{\lambda}(T)\mathrm{DEM(}T\mathrm{)}\mathrm{d}T,
	\end{equation}
	where $K_{\lambda}(T)$ denotes the response function of the corresponding filter which depends on elemental abundances and the temperature of the emitting plasma as well as on the sensor sensitivity \citep{HK2012}. 
	We assumed photospheric element abundances, as up to the lower corona, CHs show abundances close to photospheric values \citep{feldman1998,feldman2003}. For calculating the instrument response, abundances were taken from the CHIANTI 9 database \citep{chianti97,chianti19} and the filter response function from the \textsf{aia\_getresponse.pro} in the SolarSoftware (SSWIDL) package. To infer the DEM from \textit{AIA} data, we applied the regularized inversion technique developed by \cite{HK2012} to reconstruct the DEM curve for each binned pixel from the six coronal \textit{AIA} EUV channels. The temperature range considered for DEM analysis was chosen from 0.2 to 5\,MK. Integration of the DEM over the temperature range yields the total emission measure 
	\begin{equation}
	\label{Eq-EM-def}
		\mathrm{EM} = \int_{T} \mathrm{DEM(}T\mathrm{)}\mathrm{d}T.
	\end{equation}
	Following \cite{cheng2012} and \cite{vanninathan2015}, the mean temperature can be estimated by the emission weighted temperature
	\begin{equation}
	\label{Eq-T-def}
		\overline{T} =\frac{\int_{T} \mathrm{DEM(}T\mathrm{)}T\mathrm{d}T}{\mathrm{EM}}.
	\end{equation}
	Assuming a filling factor of unity, the plasma density can then be derived from Equation \ref{Eq-DEM-def} as
	\begin{equation}
	\label{Eq-n-def}
		\overline{n} =\sqrt{\frac{\text{EM}}{h}},
	\end{equation}
	with $h$ the column height of emitting plasma along the LOS.
	As an estimate for the column height of plasma contributing to the measured emission, we use the hydrostatic scale height 
	\begin{equation}
	\label{Eq-h-def}
		h =\frac{\mathrm{k_{B}}T}{\mu \mathrm{m_{H}} g}.
	\end{equation}
	Using $\mu = 0.64$ and $\mathrm{m_{H}} = 1.67\times10^{-27}$ kg for a fully-ionized solar plasma and assuming a typical CH plasma temperature around 0.9\,MK, one arrives at a typical scale height $h \approx 42$~Mm.
	   
	\section{Effect of PSF Deconvolution in Reconstructed DEMs}
	\label{sect:Deconv}
	\cite{wendel2018} pointed out that any high-temperature component in the CH plasma distribution inferred by spectroscopic analysis is likely an artifact due to instrument scattered light. In order to reduce the effect of instrument stray light on the recovered DEM, we deconvolved the data with the instrument PSF provided by the \textsf{aia\_calc\_psf.pro} routine. This is in particular relevant for studying regions of low emission such as CHs. To estimate the effectiveness of the deconvolution in reducing stray light over low-emission regions, we analyze the intensity profiles over areas occulted by the Venus transit on 6 June 2012 and a lunar eclipse on 6 August 2013.\par
\begin{figure}[t]   
	\centering
	\includegraphics[width=1.0\textwidth,clip=]{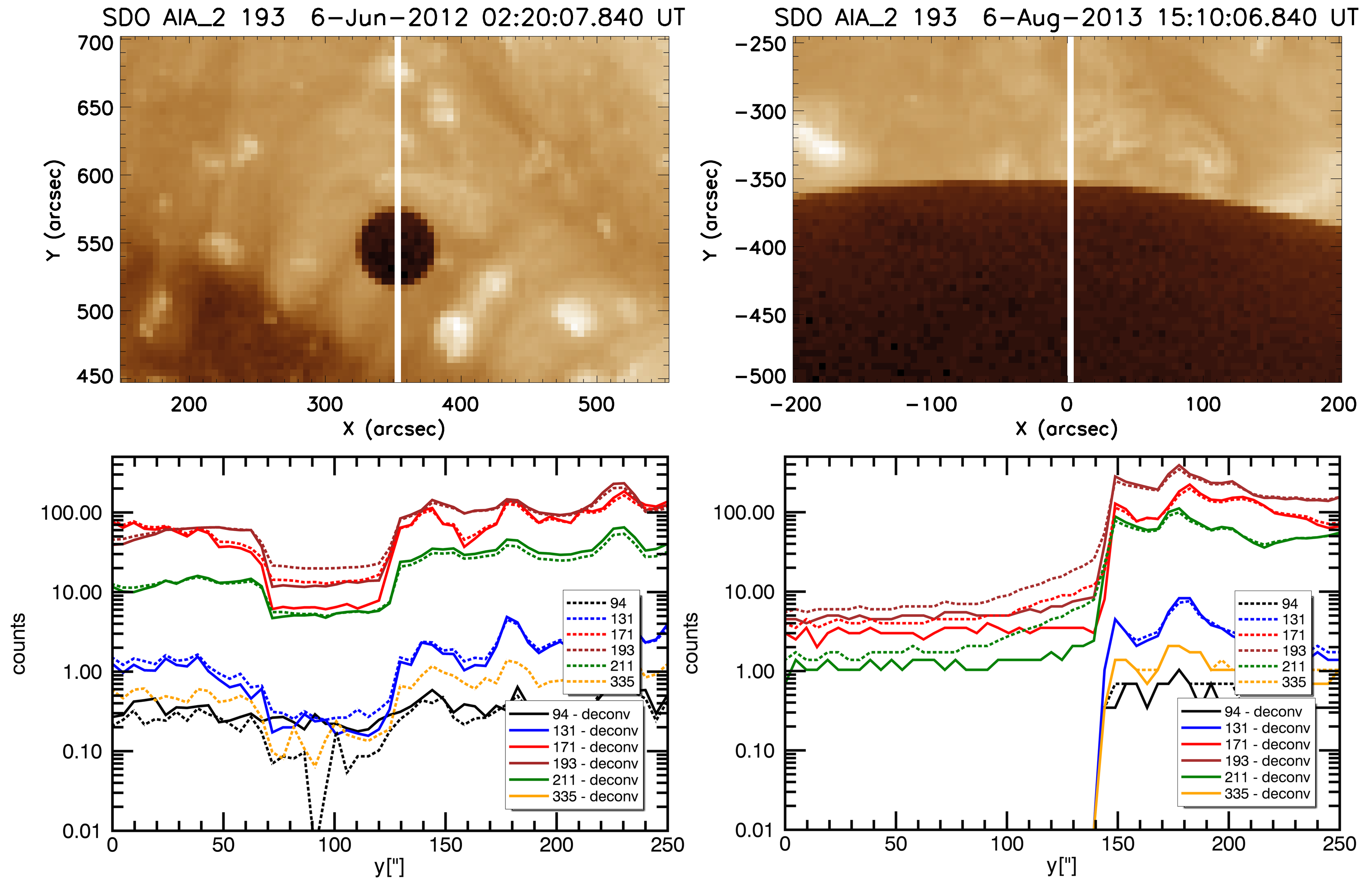}
	\caption{\textit{Top}: \textit{AIA} 193\,\AA \ images of the Venus transit (\textit{left}) and lunar eclipse (\textit{right}). The \textit{white vertical} lines indicate the regions, from where the intensity profiles shown in the bottom panels were derived. \textit{Bottom}: Comparison of intensity profiles before and after deconvolution from the PSF for low-emission areas. \textit{Left}: Intensity profile for the \textit{AIA} EUV channels along the \textit{indicated white line} during the 2012 Venus transit. \textit{Right}: Profile during a lunar eclipse.}
	\label{F-deconvProfiles}
	\end{figure}

	Figure \ref{F-deconvProfiles} (bottom panels) compares the intensities of data processed solely with \textsf{aia\_prep.pro} (level 1.5) with the intensities of the deconvolved data (level 1.6) along white lines indicated in the images in the top panels. Over the areas shaded by Venus (left) or the Moon (right), deconvolution reduces the remaining intensity for most channels, but some DNs still remain owing to long-range scattered light due to the microroughness of the telescope mirrors \citetext{private communication, M. Cheung, 2019}. The deconvolved data also show better contrast along the edge of the lunar eclipse. 
	\par
\begin{figure}[t]   
	\centering
	\includegraphics[width=1.0\textwidth,clip=]{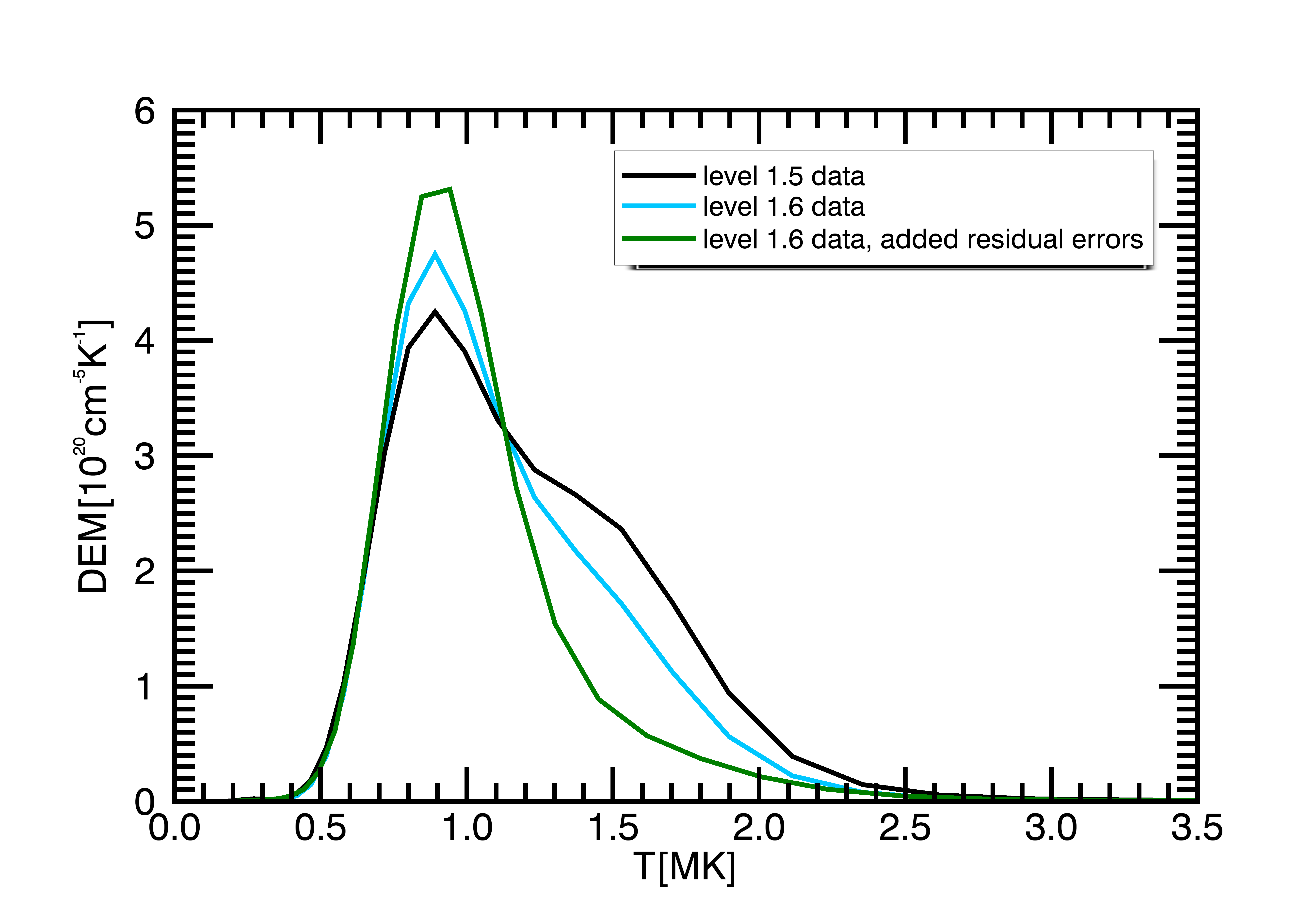}
	\caption{Comparison of the DEM averaged over the entire CH (on 15 February 2012) computed from data deconvolved with the instrument PSF (level 1.6, \textit{blue}) and without deconvolution (level 1.5, \textit{black}). The high-temperature tail is significantly reduced by the deconvolution as a result of the stray-light correction. Adding the residual DNs determined for the 2012 Venus transit to the uncertainties in the data further reduces the high-temperature emission (\textit{green}).}
	\label{F-deconvDEMComparison}
	\end{figure}

	To quantify the significance of the remaining stray light for analyzing CHs, Table \ref{T-deconv} shows the remaining DNs after deconvolving with the instrument PSF for typical Venus and lunar pixels compared with the mean DNs of the CH under study on 6 May 2012 for each \textit{AIA} channel used. Comparing Venus and CH, for the 131\,\AA \ and 171\,\AA~channels, the remaining intensity is small compared to typical CH intensities. For the 193\,\AA \  channel the remaining stray light is about 50\,\% the intensity of typical CH DNs. For the 211\,\AA \ and 94\,\AA \ channels, the contribution from stray light is about as high as the measured intensity inside the CH. Since the 171\,\AA \ and 131\,\AA \ filters, which sample cooler plasma inside the CH (Figure \ref{F-AIAFilterResponse}), are less affected by stray light than the filters sensitive to hotter plasma (Table \ref{T-deconv}), we conclude that most of the high-temperature emission in the recovered DEM of CHs is likely due to contamination from regions outside the CH.
	\begin{center}
		\begin{table}[ht]
			\caption{Remaining DNs after PSF deconvolution for typical Venus and lunar pixels compared to the mean DNs for a CH for each used \textit{AIA} channel.}
			\label{T-deconv}
			\begin{tabular}{llll||llll}
				Channel & Venus & Lunar & CH   & Channel & Venus & Lunar & CH   \\ \hline
				94\,\AA  & 0.2   & 0.0   & 0.3  & 193\,\AA & 12.4  & 5.1   & 23.5 \\
				131\,\AA & 0.2   & 0.0   & 1.2  & 211\,\AA & 5.3   & 1.2   & 6.2  \\
				171\,\AA & 6.4   & 3.4   & 67.9 & 335\,\AA & 0.0   & 0.0   & 0.3  \\ 
			\end{tabular}
		\end{table}
	\end{center}
	Figure \ref{F-deconvDEMComparison} shows the effect of deconvolving the data with the PSF and accounting for residual stray light on the recovered DEM of CHs. It compares the average DEM for an entire CH computed with the original, \textit{i.e.} not-PSF-deconvolved level 1.5 data (red), with the DEM where the PSF was taken into account (blue), and with the DEM where we have additionally increased the presumed errors in each \textit{AIA} channel due to long-range scattered stray light (green). Thereby, we have used the residual DNs of the Venus transit from Table \ref{T-deconv} as an estimate on the long-range scattered light and add these to the error budget of the \textit{AIA} channels given by the SSWIDL routine \textsf{aia\_bp\_estimate\_error.pro}. We assume this to be a reasonable estimate, as the eclipse happened during the lifetime of the CH under study, so that the EUV emission from the solar disk and thus scattered light are comparable in these data sets.	When taking the PSF into account, the high-temperature tail of the CH plasma is significantly reduced compared to the original data. Considering residual stray light as detailed above, the hot-temperature component even almost vanishes. To compute more accurate DEMs, we therefore account for both PSF and residual stray light.
	\par 
	\section{Results}
	\subsection{General Features of the CH DEM Distribution}
\begin{figure}[t]   
		\centering
		\includegraphics[width=1.0\textwidth]{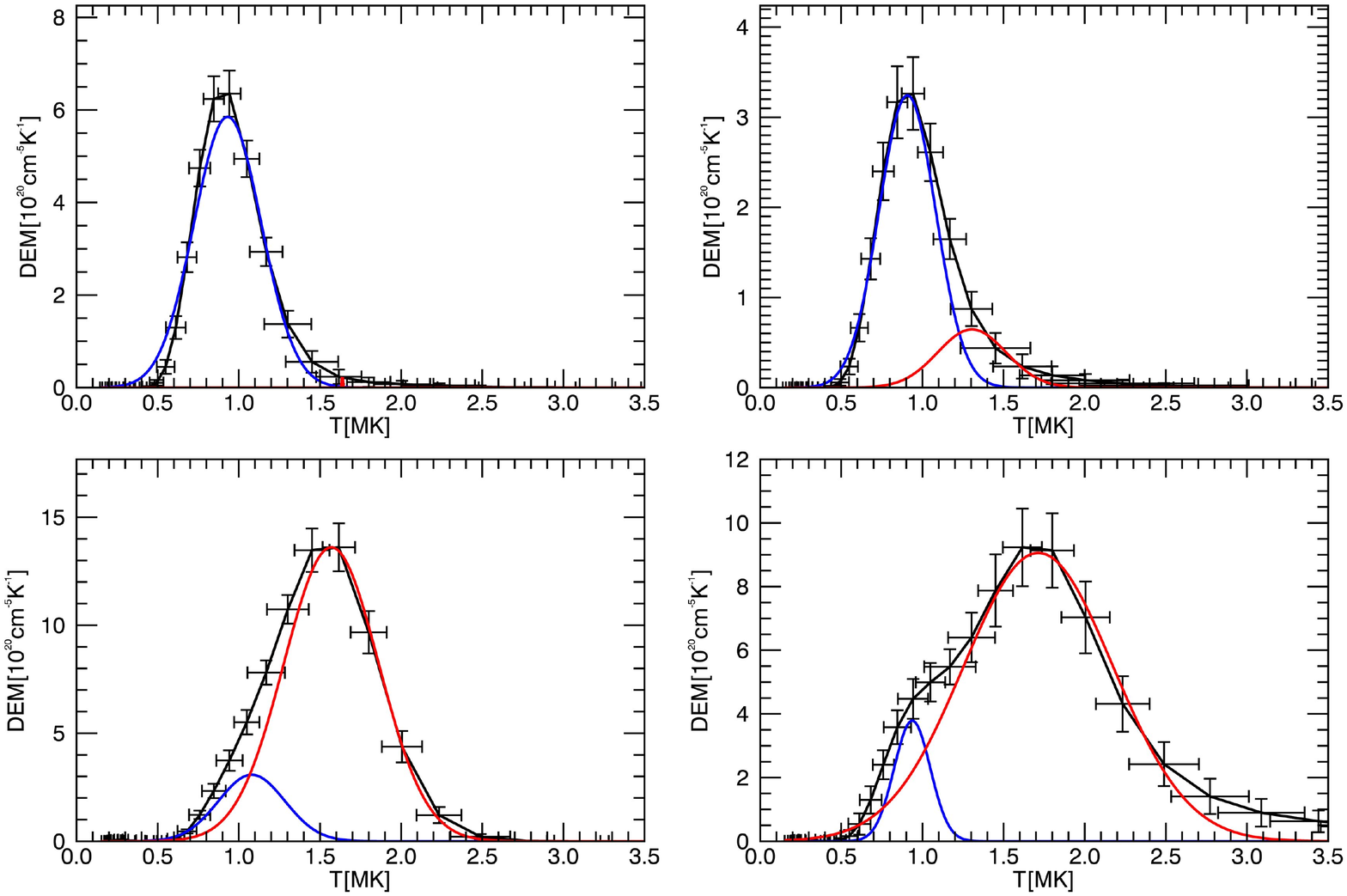}
		\caption{Example of DEM distributions for CH pixels (\textit{top row}) and quiet Sun (\textit{bottom row}) for 30 June 2012 04:00~UT.
		The \textit{red} and \textit{blue} curves are best fits to the DEM using the sum of two Gaussian components. Locations of the chosen pixels are indicated in Figure \ref{F-CH193A}.}
		\label{F-DEMComp}
	\end{figure}

	Over the ten solar rotations of the CH's lifetime, we derived the DEMs of each 8$\times$8 binned pixel of the \textit{AIA} images, when the CH was close to the central meridian.
	Figure \ref{F-DEMComp} shows examples of the derived DEM distributions for CH (top panels) and quiet Sun (QS) pixels (bottom panels). The total EM for CH regions is about ten times smaller than of the QS with the shape of the DEM also differing.
	The DEM distribution for CH regions shows a maximum of the emission at a temperature of around 0.9\,MK with a small secondary contribution from emission centered between 1.5\,--\,2.0\,MK.
	DEM from QS regions exhibits peak emission between 1.5\,--\,2.0\,MK with a secondary 0.9\,--\,1.1\,MK component being sometimes present. Also shown in Figure \ref{F-DEMComp} are the results of fitting the DEMs with the sum of two Gaussian functions centered at those temperatures (solid red and blue lines). The finding that coronal plasma outside CHs also shows emission from a component centered around CH temperatures is consistent with the reasoning of \cite{landifeldm2008}, who concluded that coronal plasma may be a superposition of CH, QS, and active-region plasma, with CHs missing the hotter components.
	\par
	\par
\begin{figure}[t]   
		\centering
		\includegraphics[width=1.0\textwidth,clip=]{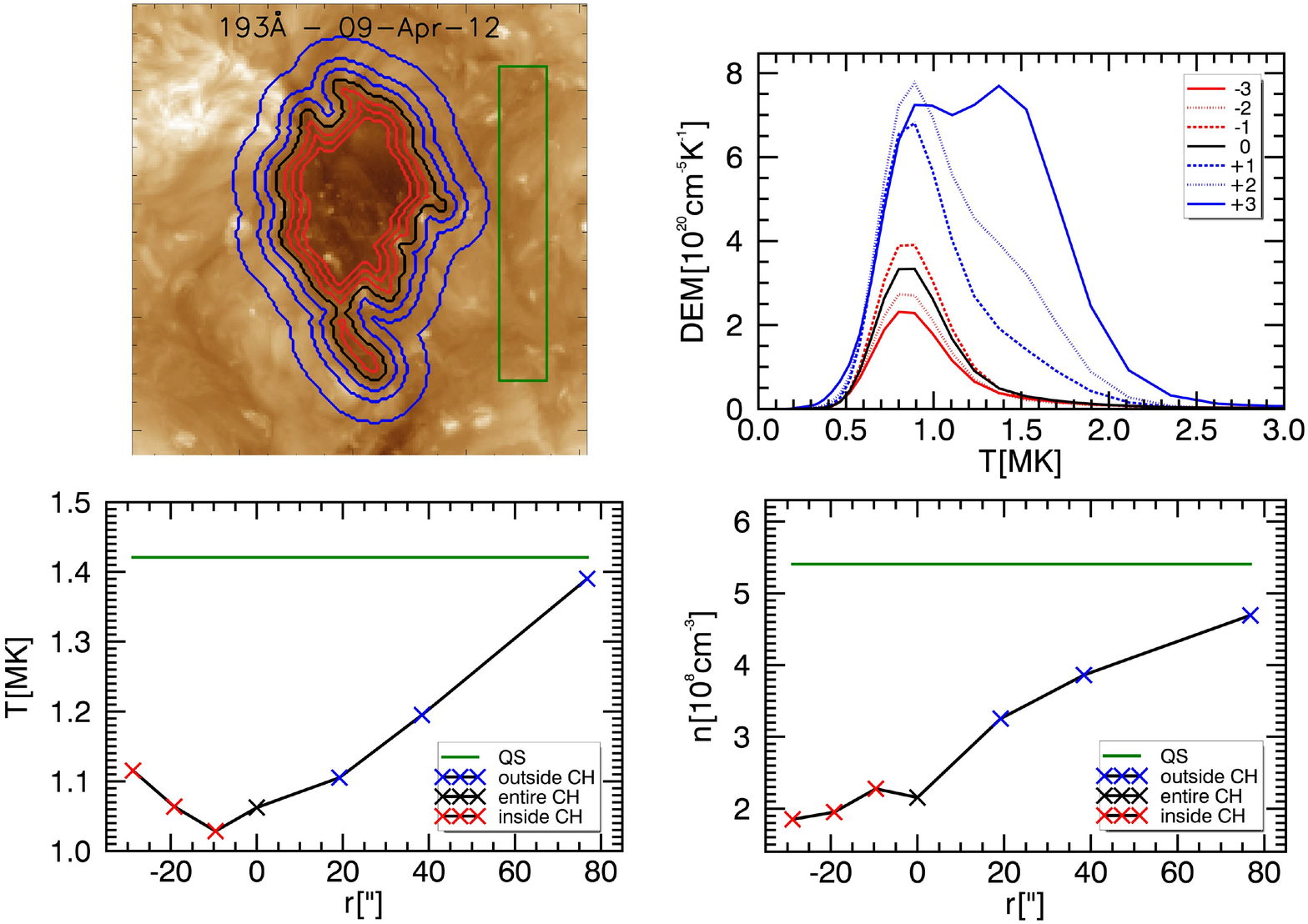}
		\caption{\textit{Top row}: \textit{AIA} 193\,\AA \ map (\textit{left panel}) with contours indicating the regions used for the  averaged DEMs shown in the \textit{right} panel. The red curves show inside, blue curves outside regions. The CH boundary is indicated in black, respectively. The distance between the contours inside the CH is \SI{9.6}{\arcsecond}. Outside the CH boundary, the distances between the contours are \SI{19.2}{\arcsecond}, \SI{19.2}{\arcsecond}, and \SI{38.4}{\arcsecond}. \textit{Bottom row}: Radial profile of DEM weighted temperature (\textit{left}) and plasma density (\textit{right}) within the indicated contours. $r=\SI{0}{\arcsecond}$ indicates the CH boundary. Shown in green are average density and temperature over the QS region indicated by the \textit{green rectangle} in the top-left panel.}
		\label{F-DEMContours}
	\end{figure}

	We further investigated the dependence of the DEM on the distance from the CH boundaries. To this aim, we separated the CH and its immediate surroundings into several segments outward and inward of the CH boundary. Inside the CH, each segment has a width of \SI{9.6}{\arcsecond}. For the first region we considered pixels that were just inside the determined CH boundary up to a distance of \SI{9.6}{\arcsecond}. For the second we used pixels at least \SI{9.6}{\arcsecond} and at most \SI{19.2}{\arcsecond} from the boundary, and so on. For the regions outside the CH boundary shown in black, the widths are \SI{19.2}{\arcsecond}, \SI{19.2}{\arcsecond}, and \SI{38.4}{\arcsecond}. We then took the averages of the DEM for all pixels inside those segments. The resulting averaged DEMs are shown in Figure \ref{F-DEMContours}. 
	The segments show a sharp difference between regions inside (red) and outside (blue) the CH. The DEM for segments \textit{inside} the determined CH boundary shows a distinct peak at $T \approx 0.9$\,MK and a small high-temperature tail of constant emission most prominent around $T \approx 1.5$\,--\,$2.0$\,MK. For the segments \textit{outside} the CH boundary, the total EM is strongly increasing and the contribution from $T \approx 1.5$\,--\,$2.0$\,MK rises. However, we note that in the segment directly adjacent to the CH boundary, the $T \approx 0.9$\,MK peak is still dominant, and only outside \textgreater \SI{38.4}{\arcsecond} from the CH boundary the emission peak at $T \approx 1.5$\,--\,$2.0$\,MK contributes equally. We note that our assumption of photospheric elemental abundances is not fulfilled outside the CH region, which may affect the derived absolute values and the ratio of cool/hot emission.
	\par 
	The radial profiles for plasma density and DEM-weighted temperature (Equations \ref{Eq-T-def}, \ref{Eq-n-def}) derived from the averaged DEM curves are also shown in Figure \ref{F-DEMContours}. The plasma density decreases towards the interior of the CH and increases outside the CH boundary. The temperature inside the CH is slightly higher towards the CH center because as the total emission decreases, the constant hot tail becomes relatively more important when calculating the mean temperature from the distribution. As the temperature of the peak does not change and the tail appears to originate mostly from stray light, this increase in plasma temperature towards the CH center is likely an effect due to limitations of the available data. The temperature increases again for segments further outside from the CH boundary. For reference, the density and temperature averaged over the QS region (utilizing the same photospheric abundances) indicated by the green rectangle in the top-left panel of Figure \ref{F-DEMContours} is shown in green.
	\begin{figure}
		\centering
		\includegraphics[width=1.0\textwidth]{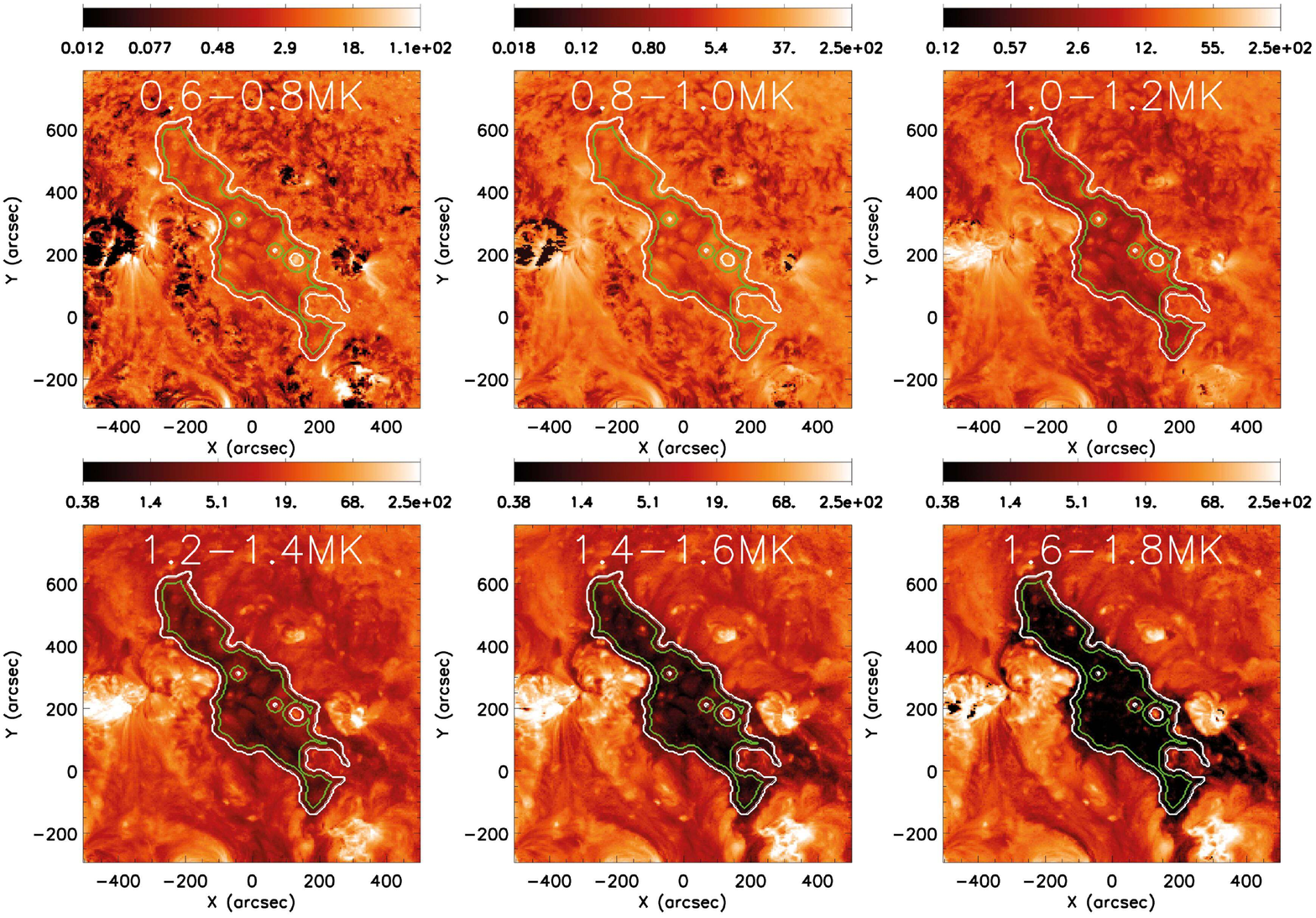}
		\caption{DEM integrated over six different temperature bins in the range 0.6 to 1.8\,MK in steps of 0.2\,MK, yielding the total emission measure (EM) in units $\mathrm{cm^{-5}}$ at those temperature ranges. Observation from 03 June 2012.}
		\label{F-EMRANGE03June}
	\end{figure}    
	As discussed previously, the favored interpretation of the non-vanishing hot ($T \approx 1.5$\,--\,$2.0$\,MK) component inside the CH seems to be stray light caused by instrumental effects and/or contribution from coronal bright points inside the CH, which show a DEM profile with a main contribution of $T > 1.5$\,MK. To reduce the influence of overlying structures and stray light from the surrounding QS, in the further analysis we limited the CH area used for the DEMs to pixels that are at least three pixels (\SI{14.4}{\arcsecond}) inside any CH boundary, in addition to deconvolving the data with the instrument PSF and accounting for the residual counts in the error budget as described in Section \ref{S-DataMethods}. Using only pixels well inside the CH has the added benefit of excluding regions where the assumption of photospheric abundances may not be valid. Figure~\ref{F-CH193A} shows these reduced CH masks overplotted in black alongside the initial CH masks derived from the image segmentation in white.
	\par
	Next we investigate the emission measure integrated over different temperature bins to illustrate the described difference in emission between CH and QS regions. Figure \ref{F-EMRANGE03June} shows the EM maps in 0.2\,MK bins over the temperature range $T=0.6$ to $1.8$\,MK for the CH on 03 June 2012. In the cooler bins up to 1.2\,MK, the CH does not appear as a distinct region of reduced emission when compared to its surroundings. In contrast, in the EM bins at $T > $ 1.2\,MK, the CH region clearly sticks out as a region of strongly reduced emission compared to the ambient corona. This finding holds for all solar rotations of the CHs lifetime (see Appendix \ref{App-EM}).
	
	\subsection{Evolution of Plasma Parameters over the CH Lifetime}
\begin{figure}[t]   
		\centering
		\includegraphics[width=1.0\textwidth]{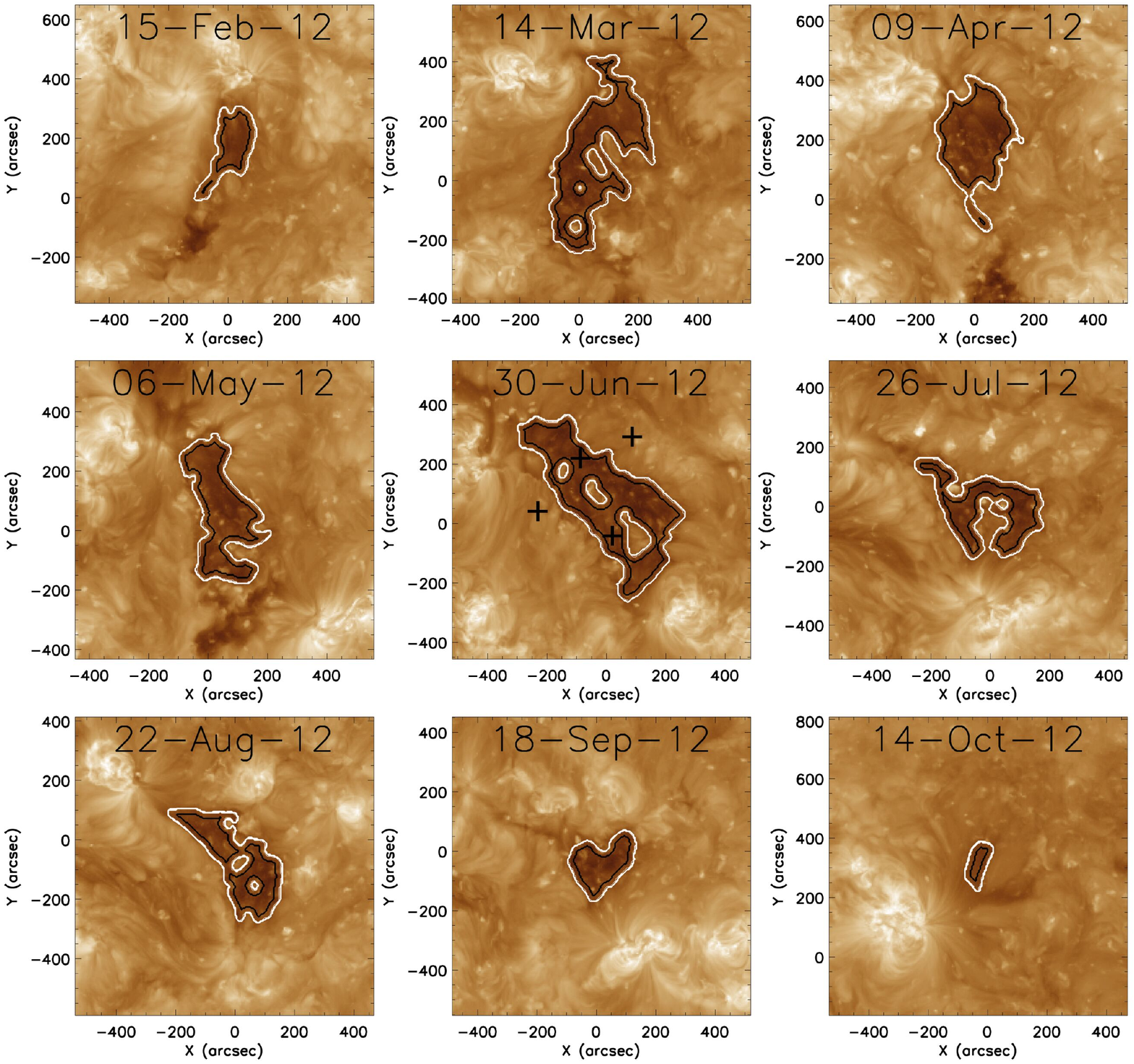}
		\caption{\textit{AIA} 193\,\AA \ filtergrams for nine solar rotations during the lifetime of the CH under study. The CH boundaries inferred from the \textit{AIA} 193\,\AA \ intensity are \textit{shown in white}. \textit{Black lines} show regions three pixels inside from the indicated CH boundary. Pixels for which we show the DEM in Figure \ref{F-DEMComp} are indicated by \textit{black crosses}.}
        \label{F-CH193A}
	\end{figure}

\begin{figure}[t]   
		\centering
		\includegraphics[width=1.0\textwidth,clip=]{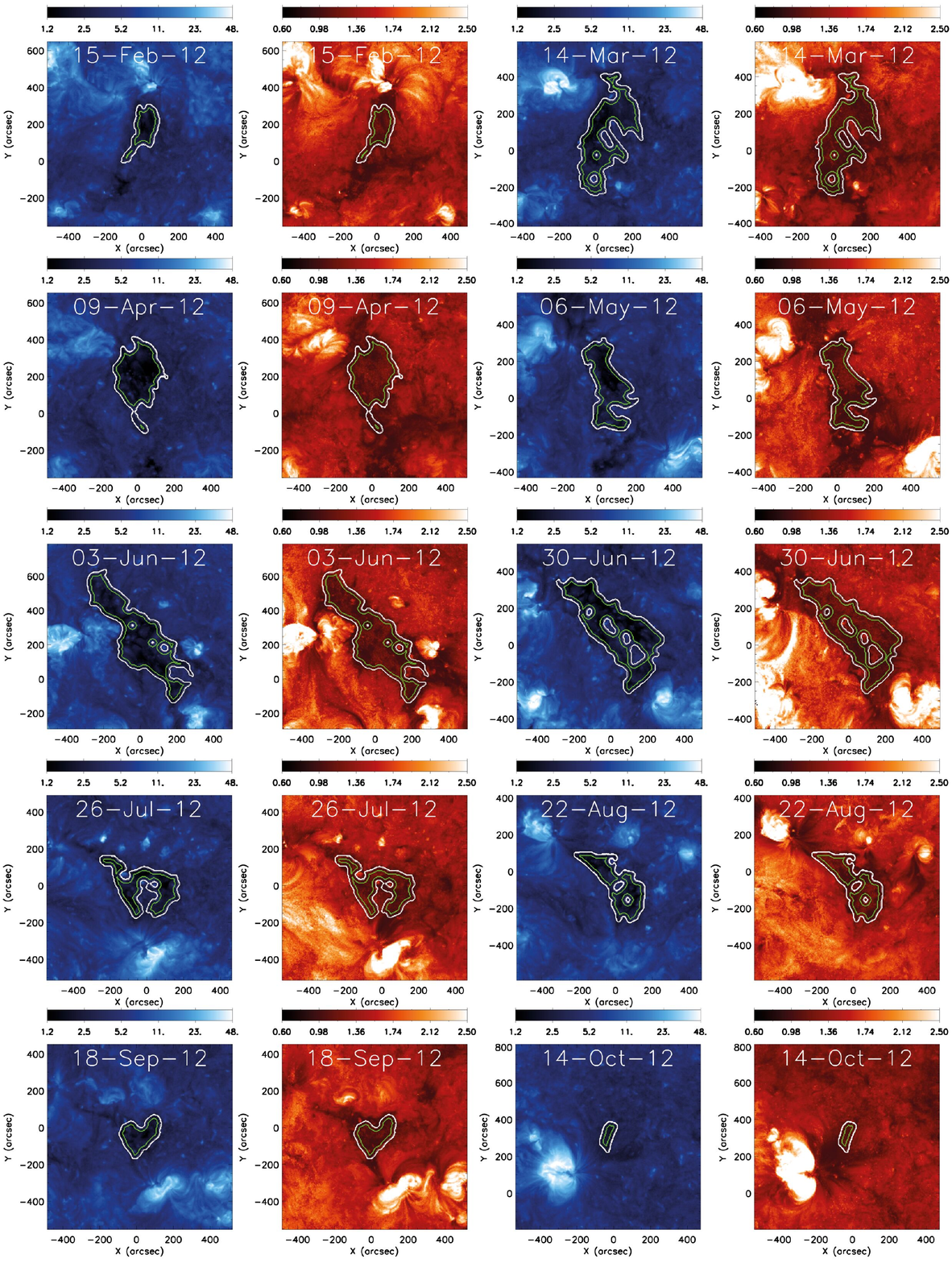}
		\caption{Maps of the mean plasma density [$\mathrm{cm^{-3}}$] (Equation \ref{Eq-n-def}) in \textit{blue} and mean temperatures [MK] (Equation \ref{Eq-T-def}) in \textit{red} together with the initial (\textit{white}) and reduced (\textit{green}) CH boundaries for all ten solar rotations where the CH was observed.}
		\label{F-meanNT}
	\end{figure}

	Figure \ref{F-CH193A} shows snapshots of the CH under study in the \textit{AIA} 193\,\AA \ filter for nine solar rotations from 15 February 2012 to 13 October 2012 illustrating the CH evolution with a growth of the CH's area followed by a period of maximal area and a decaying phase of diminishing area. CH boundaries used by \cite{heinemann2018A,heinemann2018B} are indicated in white.
	Figure \ref{F-meanNT} shows maps of the mean density and mean ion temperature (Equations \ref{Eq-T-def} and \ref{Eq-n-def}) during the CH evolution. To quantify the changes of the DEM during the CH's lifetime, we derive the mean of all of the DEM curves of pixels inside the CH at a given solar rotation, utilizing the smaller boundaries indicated in black in Figure \ref{F-CH193A}. \par
\begin{figure}[t]
		\centering
		\includegraphics[width=0.9\textwidth,clip=]{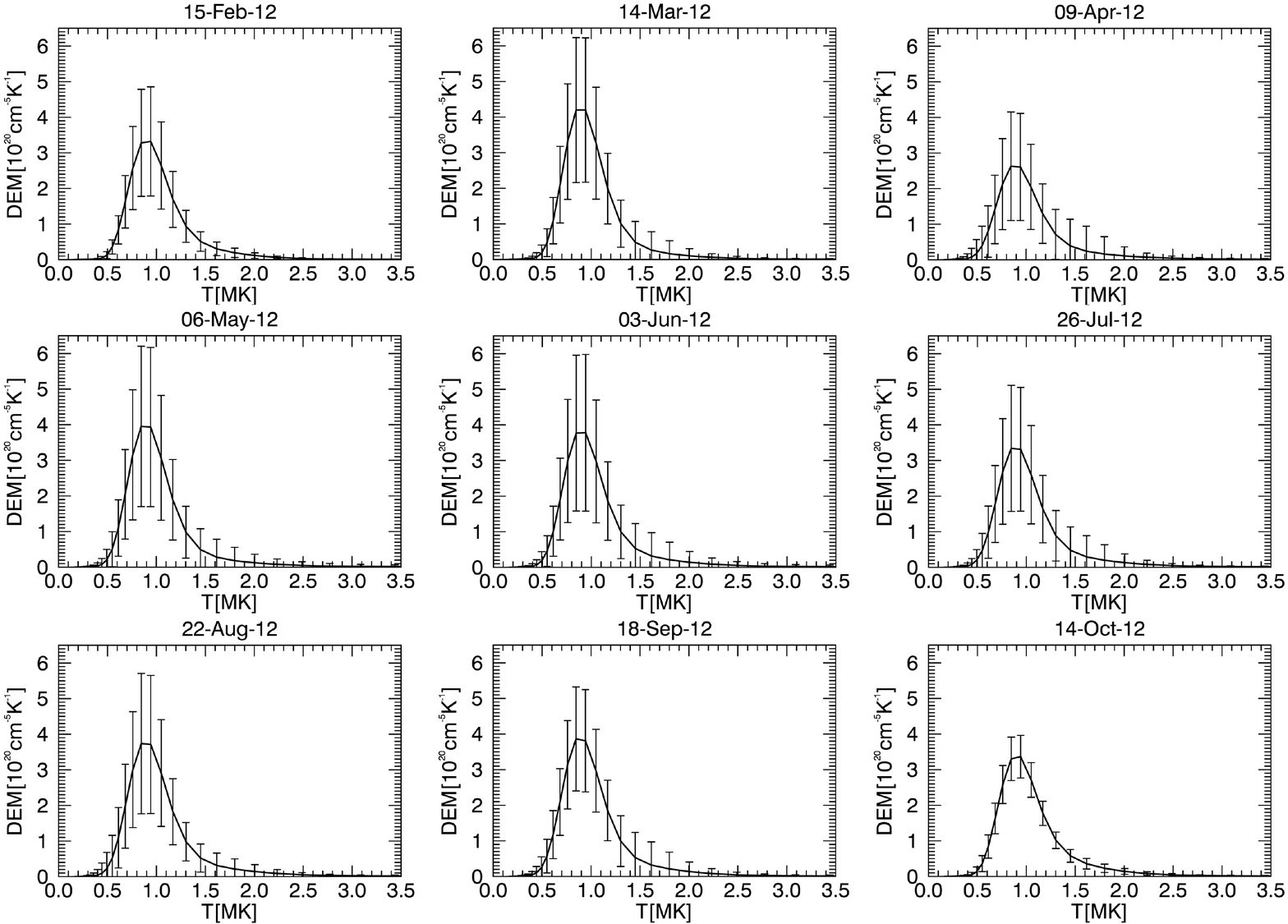}
		\caption{DEM curves averaged over all pixels inside the CH. Note that we used only pixels \SI{14.4}{\arcsecond} away from the initial CH boundary. The evolution over nine solar rotations during the CH lifetime is shown. \textit{Vertical bars} indicate the 1$\sigma$ range.}
		\label{F-avgDEM}
	\end{figure}

 Figure \ref{F-avgDEM} shows the derived mean DEMs together with the 1$\sigma$ range of values for the nine solar rotations, shown in Figure \ref{F-CH193A}. It is noted that the DEM in April 2012 is notably reduced. The reason for this steep reduction is unclear but it seems to be a real effect. We tested the DEM of the CH for different days during this rotation, and all of them showed the same behaviour.
	\par
\begin{figure}
		\centering
		\includegraphics[width=0.9\textwidth,clip=]{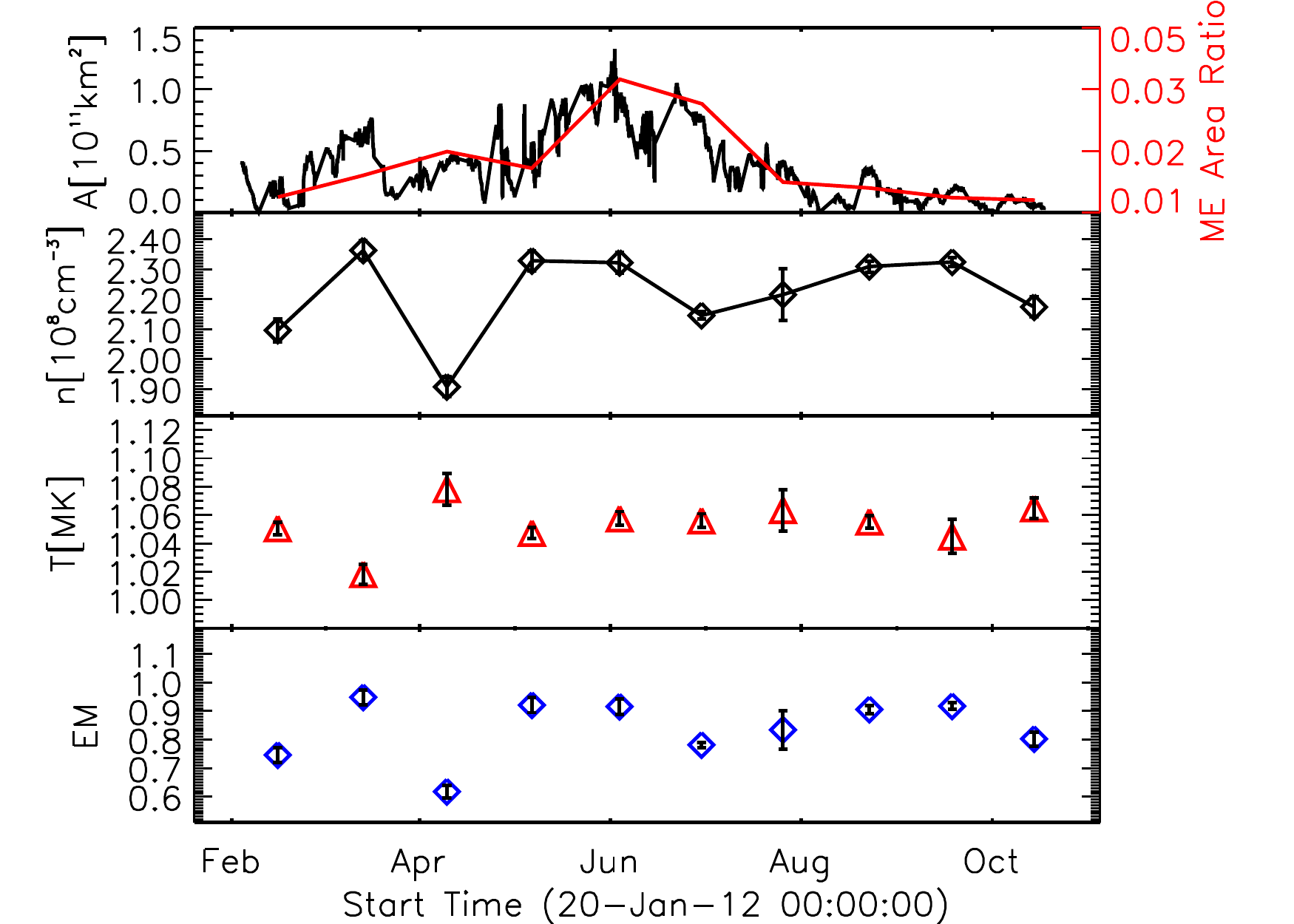}
		\caption{\textit{Top}: Evolution of the CH area (\textit{black}) and fraction of the CH area covered by magnetic elements \textgreater 50\,G (\textit{red}) from \cite{heinemann2018A, heinemann2018B}. \textit{Second and third from top}: Density and temperature evolution (Equations \ref{Eq-T-def},\ref{Eq-n-def}). \textit{Fourth panel}: Total EM (Equation \ref{Eq-EM-def}). Error bars show standard deviations for each solar rotation.}
		\label{F-Area-EM}
	\end{figure}

	Figure \ref{F-Area-EM} shows the evolution of the total EM as well as density and averaged temperature (Equations \ref{Eq-EM-def}\,--\,\ref{Eq-n-def}) together with the CH area and fraction of the CH area covered by magnetic elements \textgreater 50\,G from \cite{heinemann2018B} over the CH lifetime. Error bars indicate standard deviations over different observations within each rotation for each solar rotation, with data points showing the mean values. EM values are normalized to the maximum of the time series. 
	Comparing the evolution of EM, density and temperature with the evolution of the CH area and its coverage by magnetic flux concentrations \textgreater~50\,G, there is no clear correlation of the derived evolution of the CH plasma parameters and the previously described phases of growing, maximum, and decaying area identified by \cite{heinemann2018A,heinemann2018B}.
	
\section{Discussion and Conclusion}
	Using the DEM code developed by \cite{HK2012}, we reconstructed mean plasma temperature and density for a long-lived, low-latitude CH over its lifetime of ten solar rotations from the six coronal \textit{AIA} EUV filters. We initially recovered a non-isothermal DEM distribution for CH regions (Figure \ref{F-DEMComp}) with the main peak centered around $T=$0.9\,MK and and a secondary smaller component in the range 1.5\,--\,2.0\,MK. A similar finding was reported in previous studies such as \cite{hahn2011} who performed DEM analysis of a polar CH using \textit{Hinode}/\textit{Extreme-Ultraviolet Imaging Spectrometer} (\textit{EIS}) data and attributed the hot contribution from 1.5 to 2\,MK to structures along the LOS, contamination from streamers or closed loops of higher temperature inside the CH itself.
	\cite{wendel2018} studied \textit{Hinode}/\textit{EIS} observations of an equatorial CH and concluded that any emission from hot plasma in CHs may be the result of a previously underestimated contribution from instrumental stray light. A study of the off-disk thermal structure for a polar CH by \cite{landi2008} using spectra from the \textit{Solar and Heliospheric Observatory} (\textit{SOHO})/\textit{Solar Ultraviolet Measurements of Emitted Radiation} (\textit{SUMER}) instrument recovered a near isothermal off-disk CH plasma with a height-dependent temperature 5.85~\textless~log($T$)~\textless~6.04~K, further supporting the assumption of a non physical origin of the hot tail of the recovered CH DEMs.
	\par
     We showed that deconvolving the data with the instrument PSF reduces the hot tail in the recovered DEM distribution of CHs (Figure \ref{F-deconvDEMComparison}). Using \textit{AIA} observations of the 2012 Venus transit and a lunar eclipse, we verified the assumption that the hot tail is not actually emitted from the CH. We found that for hot \textit{AIA} channels, even after PSF deconvolution significant DNs remain for pixels in regions where the Sun is occulted (Figure \ref{F-deconvProfiles}, Table \ref{T-deconv}). We therefore used the residual counts obtained by analyzing the radial profile of the Venus transit as an additional contribution to the errors in the \textit{AIA} measurements, thereby further reducing the hot tail in the derived CH DEM profiles (Figure \ref{F-deconvDEMComparison}).
	\par
	EM maps clearly show the CH as a region of reduced intensity for temperatures \textgreater 1.2\,MK whereas for emission from cooler temperatures, no such effect is noticeable (Figure \ref{F-EMRANGE03June}, Appendix \ref{App-EM}). The radial profile of the density and temperature shows a steep decrease in the regions adjacent to the CH boundary, with relatively flat profiles inside the CH. While the density decreases away from the boundary, the temperature slightly increases towards the CH center due to a small, constant, high-temperature emission (Figure \ref{F-DEMContours}). As the temperature of the DEM maximum remains constant, we interpret this as the result of instrumental effects that lead to the described constant higher-temperature tail.
	\par
	We do not find a distinct trend linking CH area and EM/density or plasma temperature that would correspond to the reported three-phase evolution of the long-lived CH under study. But we find that once the CH is clearly observed as a region of reduced emission in the EUV images, it already has a significantly reduced density and temperature, which do not distinctly change over the CH's lifetime. Since the magnetic properties follow the same three-phase evolution as the CH area \citep{heinemann2018B}, this indicates that the mean density and temperature evolution of the CH are not strongly governed by the changing magnetic properties and fine structure of the CH. While the magnetic field is believed to be responsible for heating coronal plasma \citep{abr2006}, the bulk of the CH plasma appears to not be significantly altered by the changes in the photospheric magnetic fine structures of the dominant coronal-hole polarity. Our finding that the plasma properties in the CH do not evolve with the area and magnetic-field strength suggests that the CH plasma properties are mostly ``set'' at the CH formation and/or that all CHs have similar plasma properties. To verify these hypotheses, a statistical study covering a large set of well observed CHs is necessary.
%
\begin{acks}
    SDO image data courtesy of NASA/SDO and the AIA, EVE, and HMI science teams. A.M. Veronig, S.G. Heinemann, M. Temmer, and K.Dissauer acknowledge the financial support by the Austrian Space Applications Programme of the Austrian Research Promotion Agency FFG (ASAP-13 859729, SWAMI and ASAP-14 865972, SSCME). The authors thank Mark Cheung (LMSAL) for the insightful discussion on the AIA instrument and PSF.
\end{acks}

\section*{Disclosure of Potential Conflicts of Interest}
    The authors declare that they have no conflicts of interest.

%
\FloatBarrier
\appendix
\section{Emission Measure Maps over the CH Lifetime}
	Emission measure (Equation \ref{Eq-EM-def}) integrated over different temperature bins, illustrating the difference in emission between CH and QS regions for the solar rotations not shown in Figure \ref{F-EMRANGE03June}.
	\label{App-EM}
	\begin{figure}[!ht]
		\centering
		\includegraphics[width=0.83\textwidth]{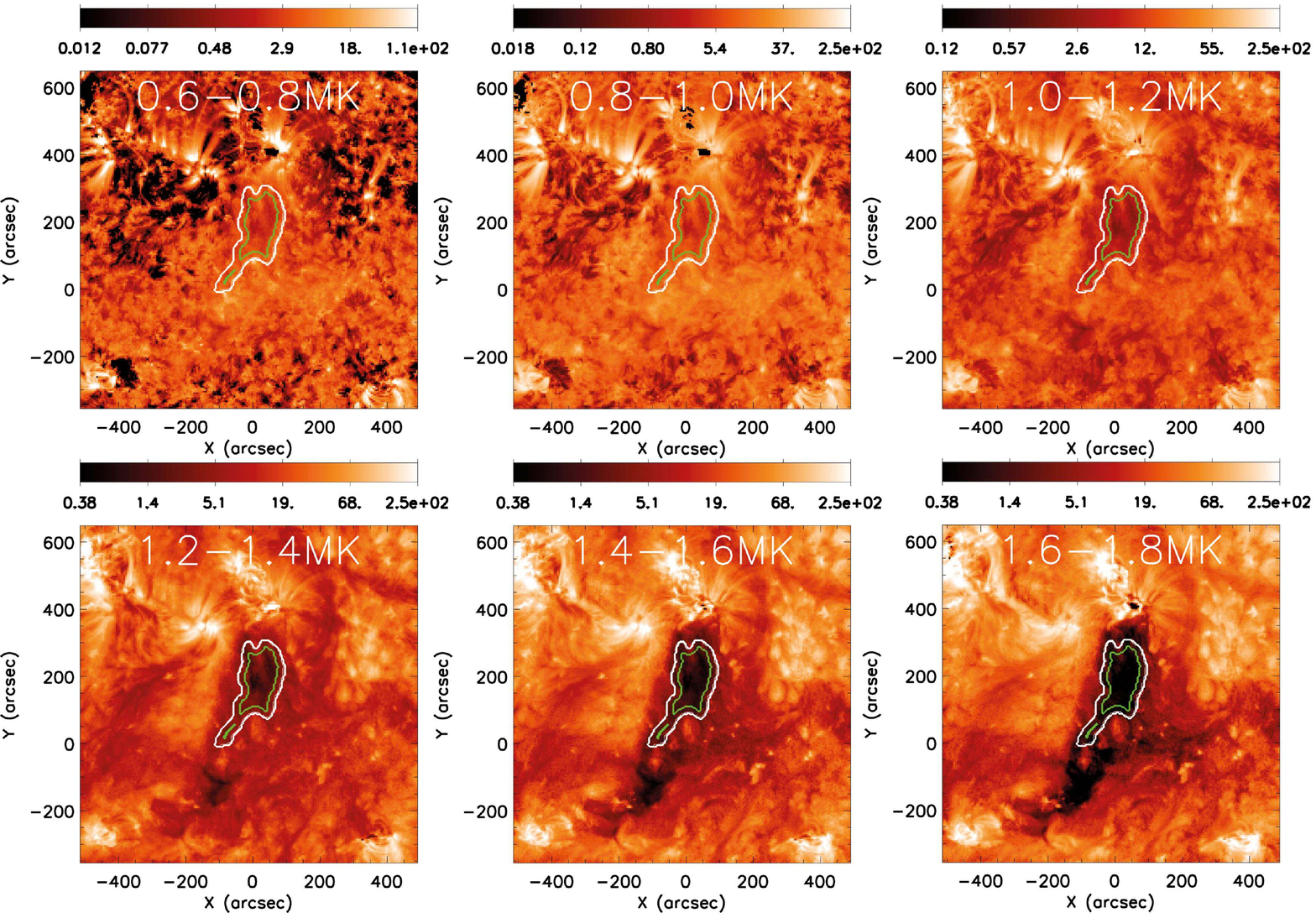}
		\caption{DEM integrated over six different temperature bins in the range 0.6 to 1.8\,MK in steps of 0.2\,MK, yielding the total emission measure (EM) in units $\mathrm{cm^{-5}}$ at those temperature ranges. \textit{White contours} indicate the initial, \textit{green} the reduced CH boundaries. Observation from 15 February 2012.}
		\label{F-EMRANGEFeb}
	\end{figure}   
	\begin{figure}[!ht]
		\centering
		\includegraphics[width=0.83\textwidth]{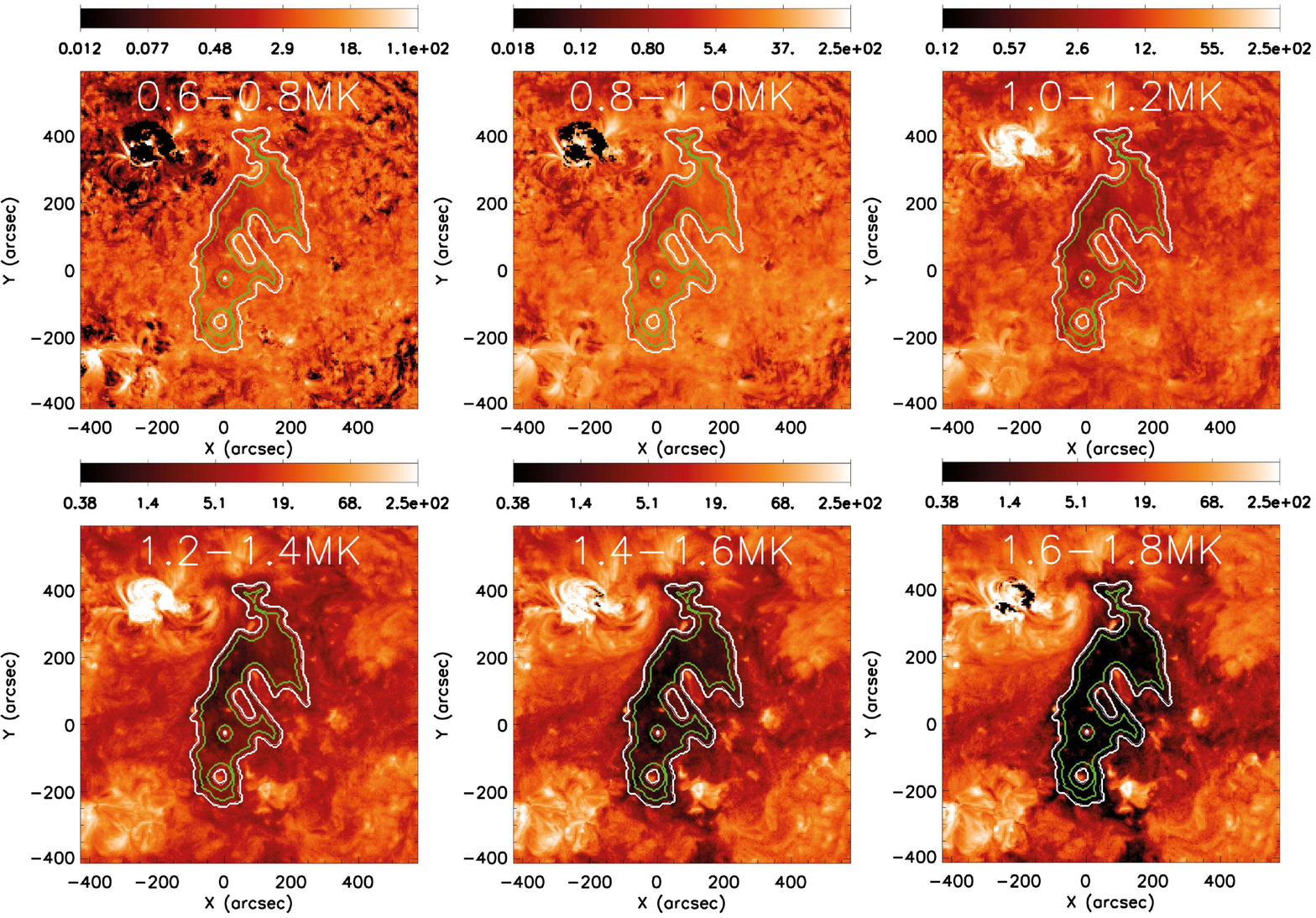}
		\caption{DEM integrated over six different temperature bins in the range 0.6 to 1.8\,MK in steps of 0.2\,MK, yielding the total emission measure (EM) in units $\mathrm{cm^{-5}}$ at those temperature ranges. \textit{White contours} indicate the initial, \textit{green} the reduced CH boundaries. Observation from 14 March 2012.}
		\label{F-EMRANGEMar}
	\end{figure}   
	\newpage
	\FloatBarrier
	\begin{figure}[!ht]
		\centering
		\includegraphics[width=0.83\textwidth]{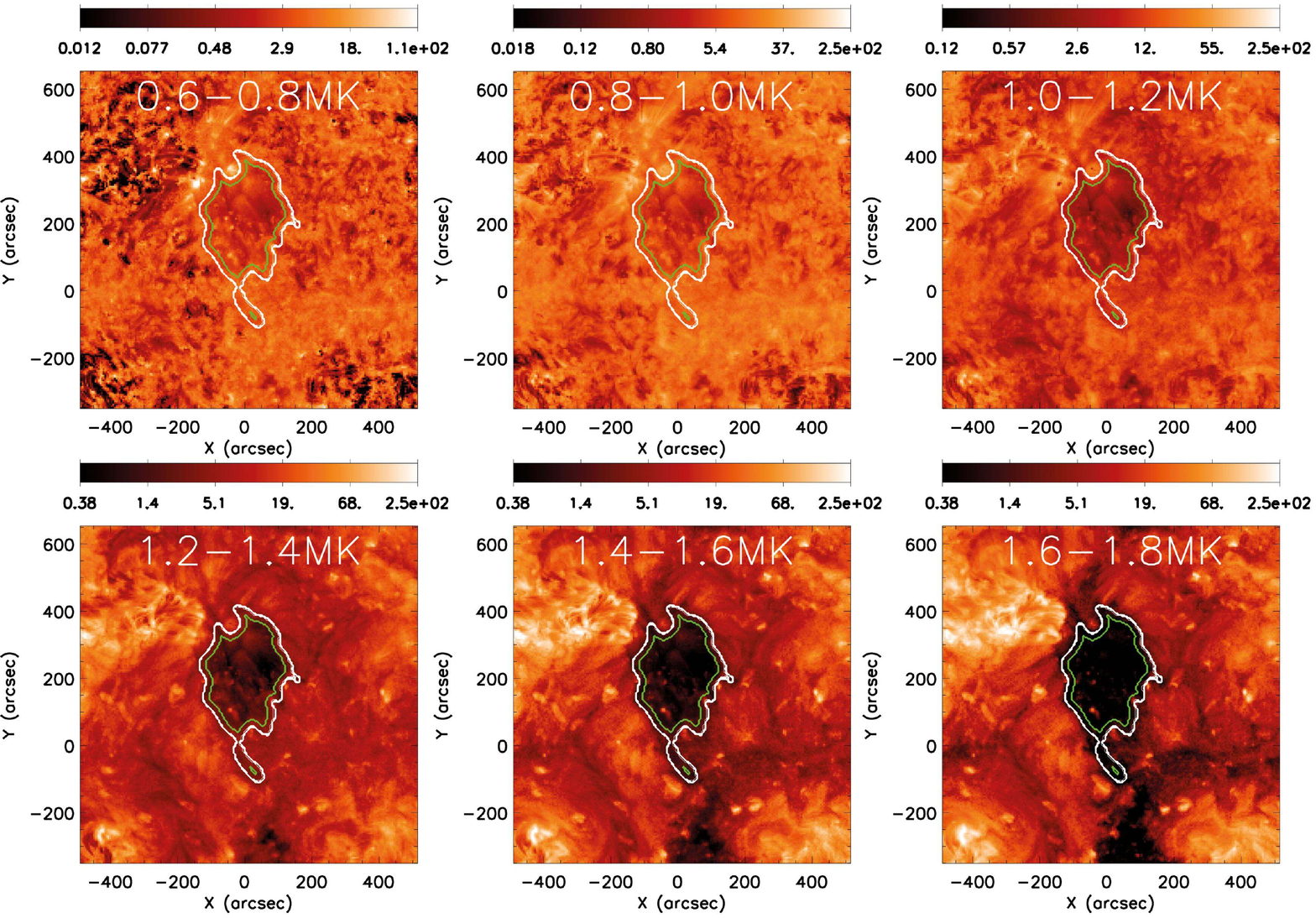}
		\caption{DEM integrated over six different temperature bins in the range 0.6 to 1.8\,MK in steps of 0.2\,MK, yielding the total emission measure (EM) in units $\mathrm{cm^{-5}}$ at those temperature ranges. \textit{White contours} indicate the initial, \textit{green} the reduced CH boundaries. Observation from 09 April 2012.}
		\label{F-EMRANGEApr}
	\end{figure}   
	\begin{figure}[!ht]
		\centering
		\includegraphics[width=0.83\textwidth]{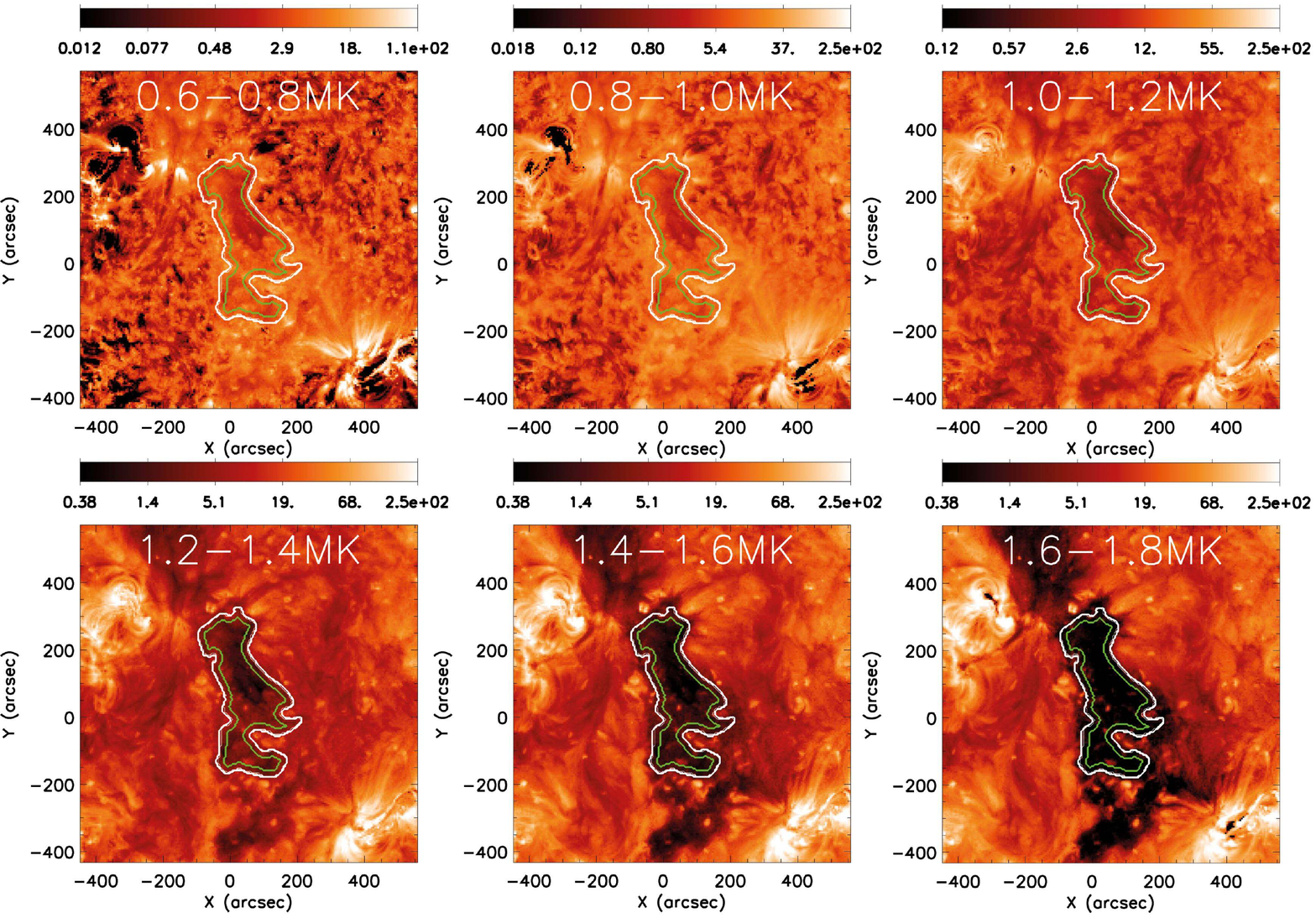}
		\caption{DEM integrated over six different temperature bins in the range 0.6 to 1.8\,MK in steps of 0.2\,MK, yielding the total emission measure (EM) in units $\mathrm{cm^{-5}}$ at those temperature ranges. \textit{White contours} indicate the initial, \textit{green} the reduced CH boundaries. Observation from 06 May 2012.}
		\label{F-EMRANGEMay}
	\end{figure}   
	\FloatBarrier
	\newpage
	\FloatBarrier
	\begin{figure}[!ht]
		\centering
		\includegraphics[width=0.83\textwidth]{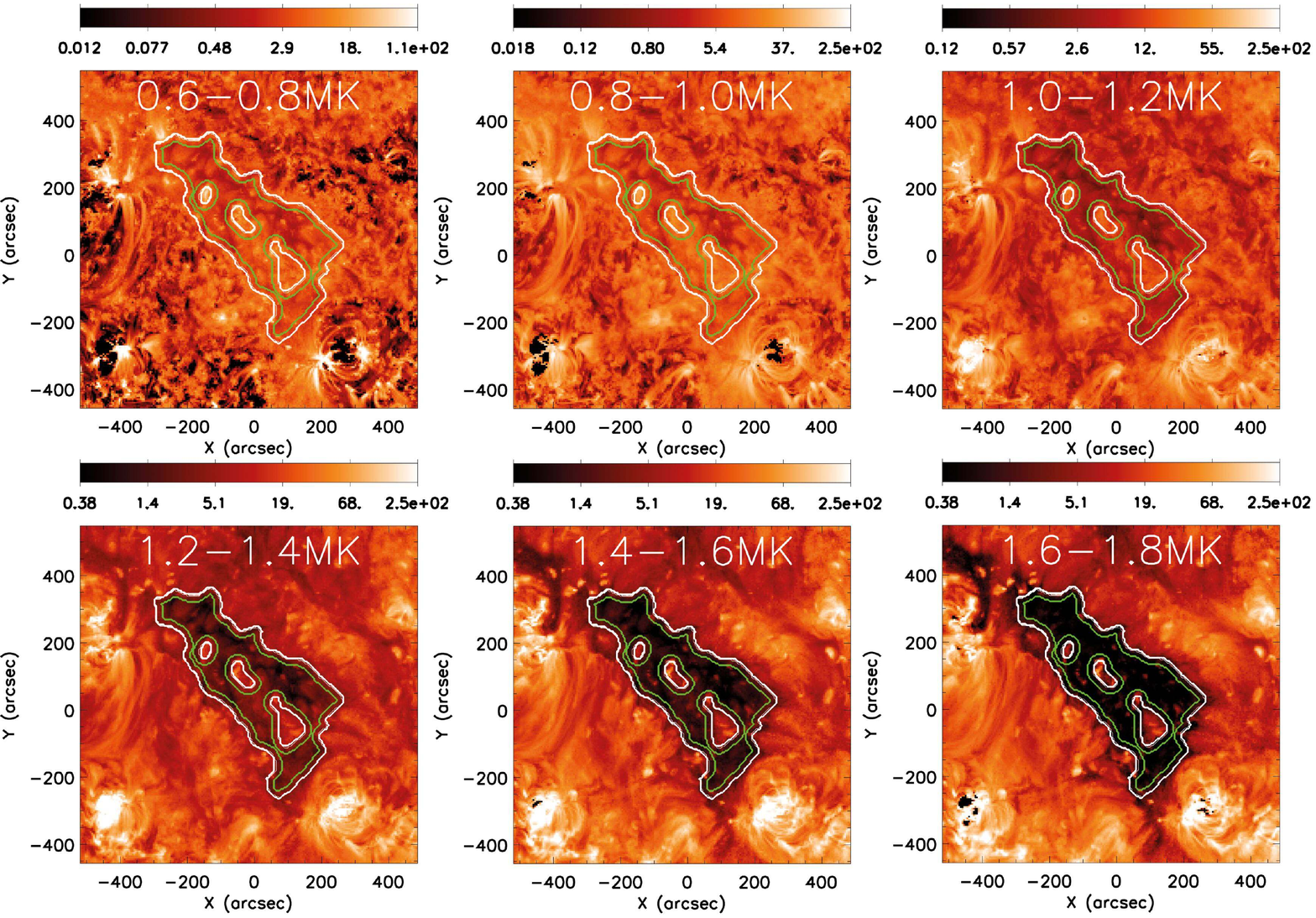}
		\caption{DEM integrated over six different temperature bins in the range 0.6 to 1.8\,MK in steps of 0.2\,MK, yielding the total emission measure (EM) in units $\mathrm{cm^{-5}}$ at those temperature ranges. \textit{White contours} indicate the initial, \textit{green} the reduced CH boundaries. Observation from 30 June 2012.}
		\label{F-EMRANGE30June}
	\end{figure}   
	\begin{figure}[!ht]
		\centering
		\includegraphics[width=0.83\textwidth]{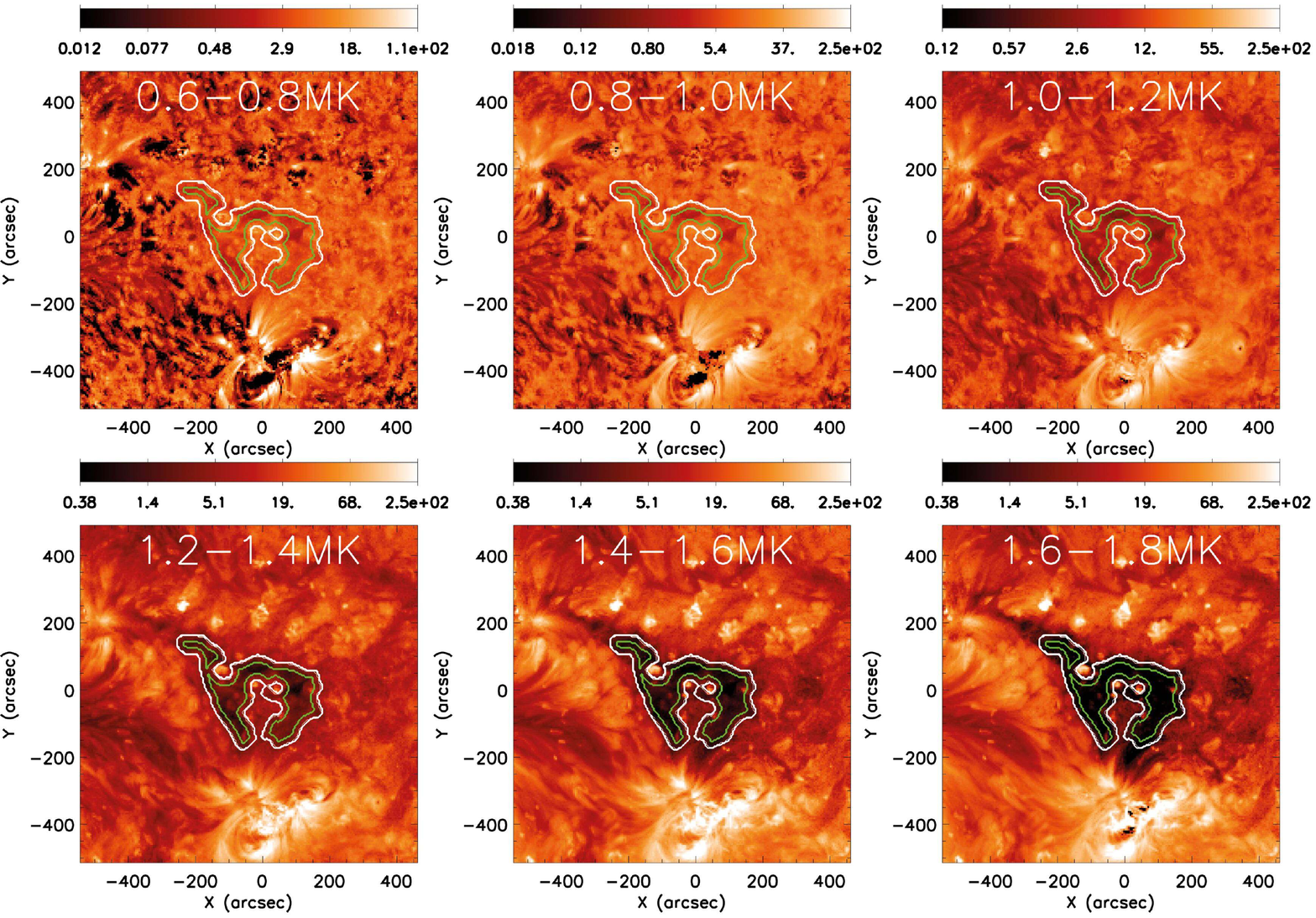}
		\caption{DEM integrated over six different temperature bins in the range 0.6 to 1.8\,MK in steps of 0.2\,MK, yielding the total emission measure (EM) in units $\mathrm{cm^{-5}}$ at those temperature ranges. \textit{White contours} indicate the initial, \textit{green} the reduced CH boundaries. Observation from 26 July 2012.}
		\label{F-EMRANGEJuly}
	\end{figure}   
	\FloatBarrier
	\newpage
	\FloatBarrier
	\begin{figure}[!ht]
		\centering
		\includegraphics[width=0.83\textwidth]{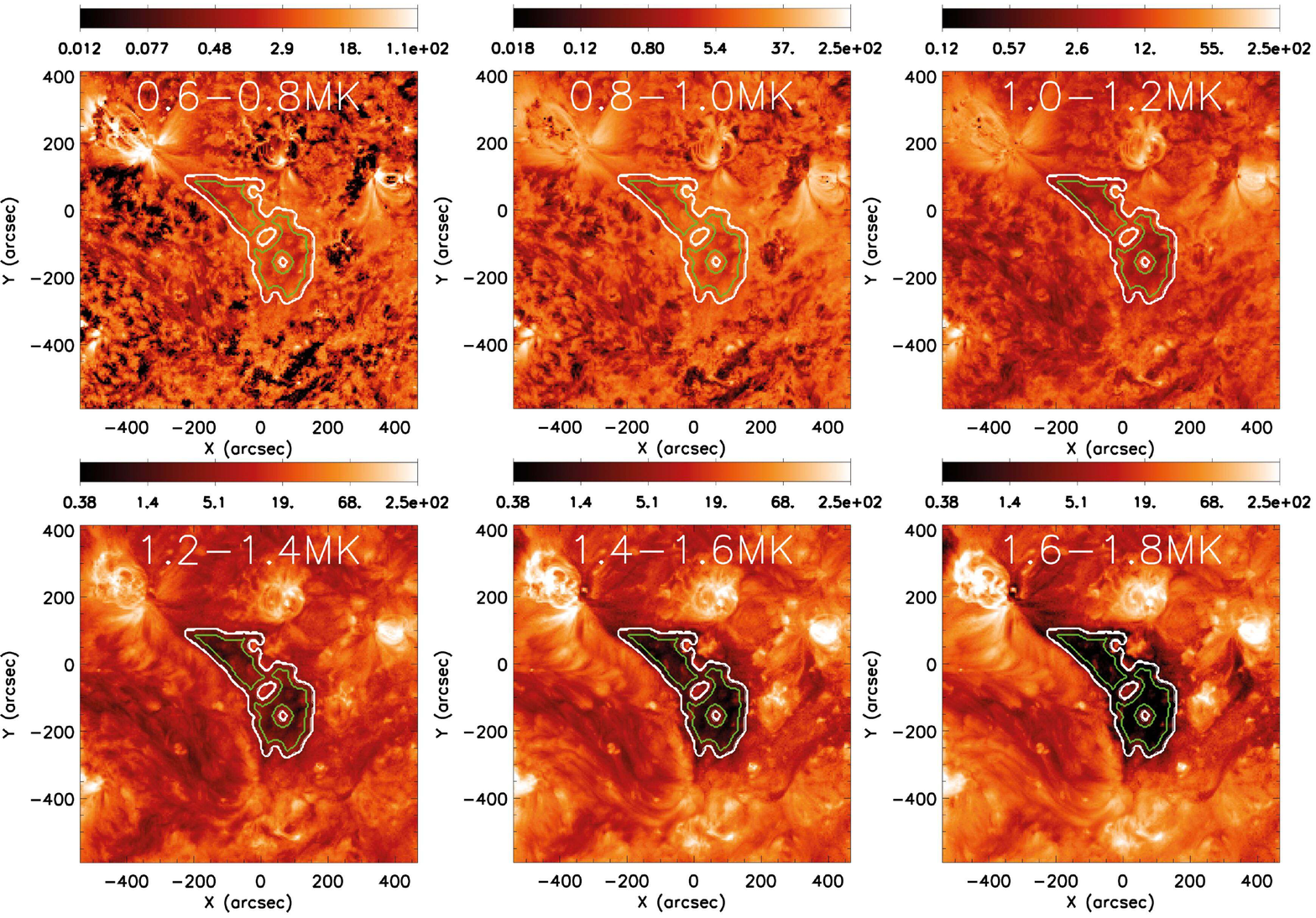}
		\caption{DEM integrated over six different temperature bins in the range 0.6 to 1.8\,MK in steps of 0.2\,MK, yielding the total emission measure (EM) in units $\mathrm{cm^{-5}}$ at those temperature ranges. \textit{White contours} indicate the initial, \textit{green} the reduced CH boundaries. Observation from 22 August 2012.}
		\label{F-EMRANGEAug}
	\end{figure}   
	\begin{figure}[!ht]
		\centering
		\includegraphics[width=0.83\textwidth]{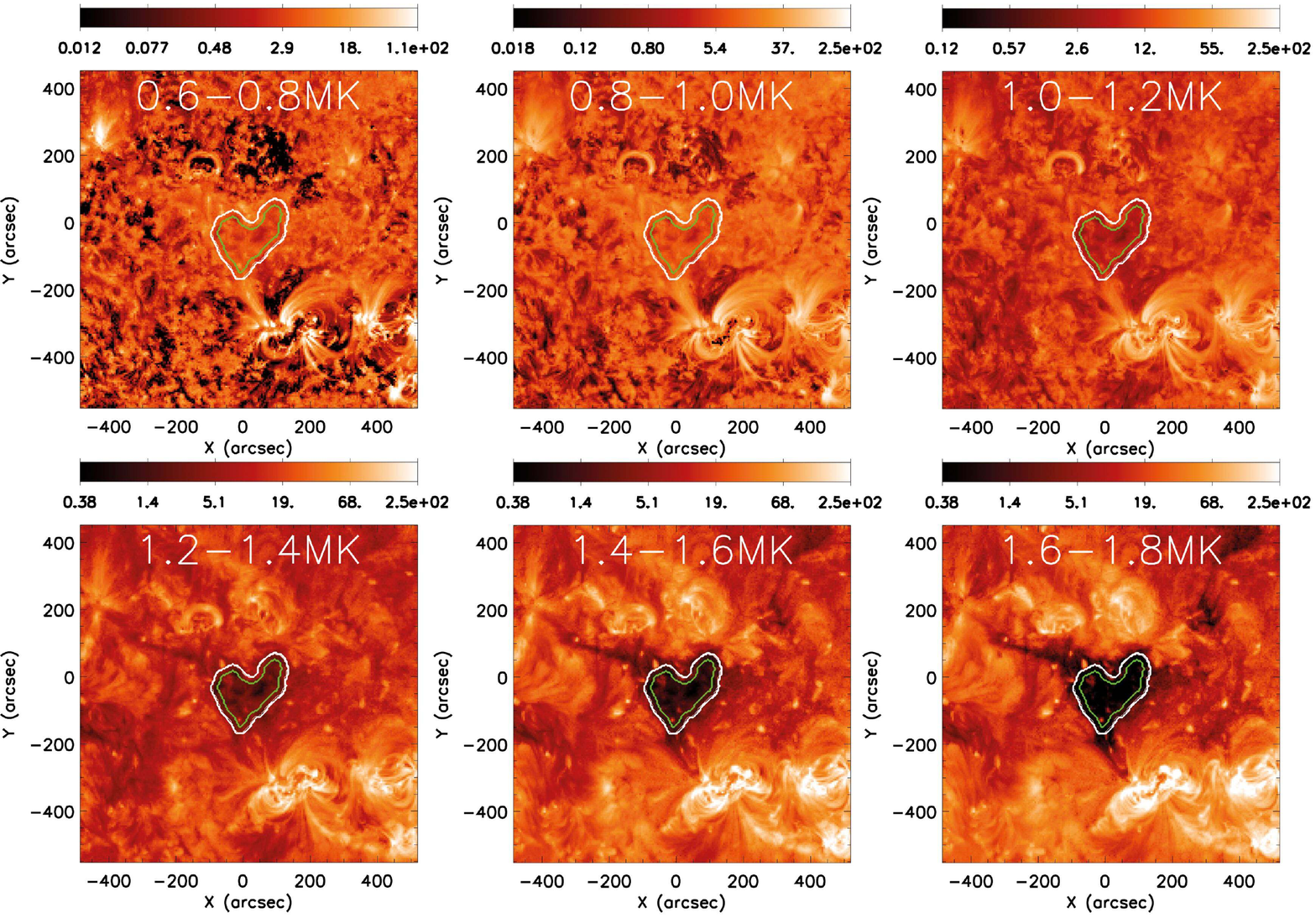}
		\caption{DEM integrated over six different temperature bins in the range 0.6 to 1.8\,MK in steps of 0.2\,MK, yielding the total emission measure (EM) in units $\mathrm{cm^{-5}}$ at those temperature ranges. \textit{White contours} indicate the initial, \textit{green} the reduced CH boundaries. Observation from 18 September 2012.}
		\label{F-EMRANGESep}
	\end{figure}   
	\FloatBarrier
	\newpage
	\FloatBarrier
	\begin{figure}[!ht]
		\centering
		\includegraphics[width=0.83\textwidth]{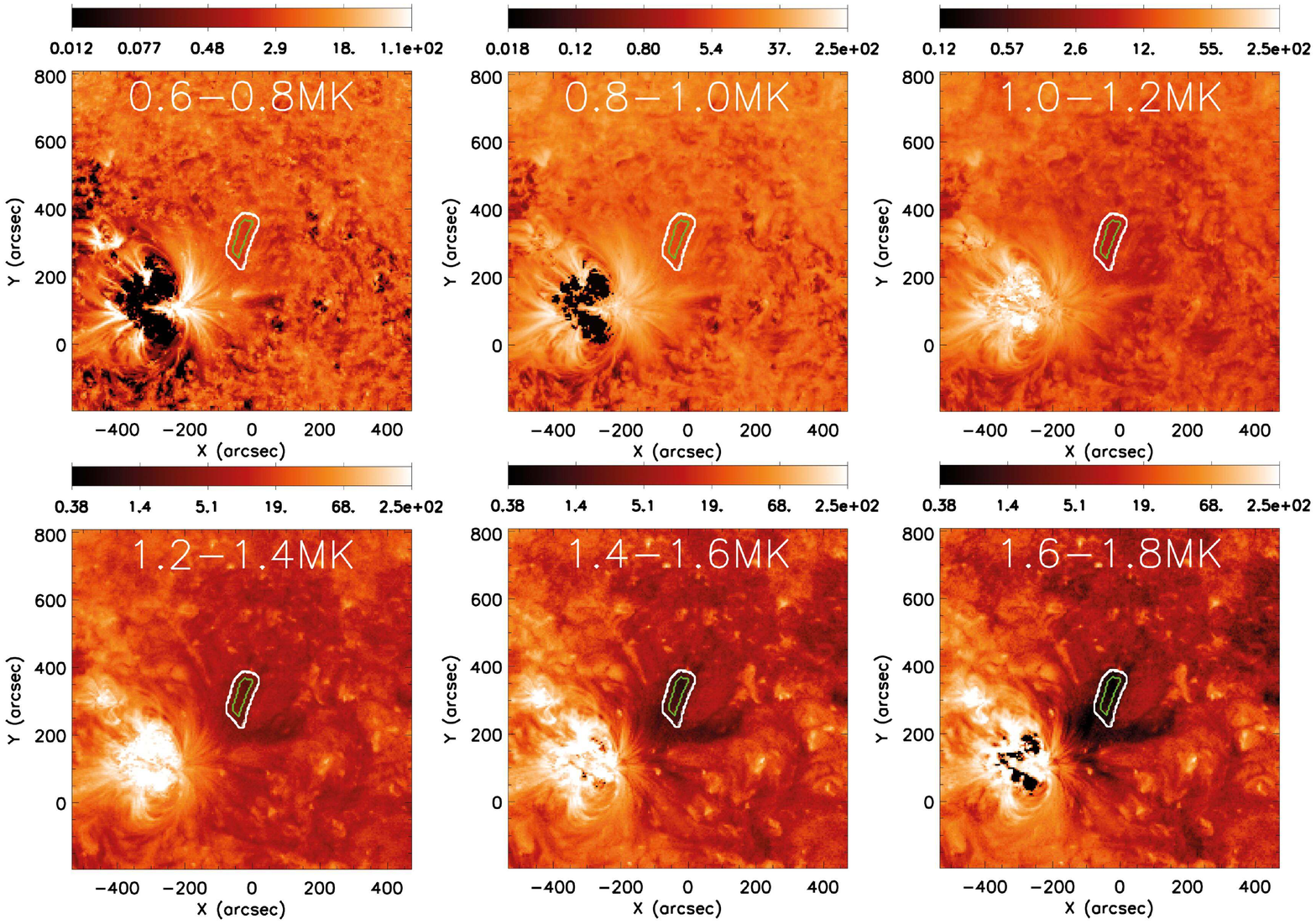}
		\caption{DEM integrated over six different temperature bins in the range 0.6 to 1.8\,MK in steps of 0.2\,MK, yielding the total emission measure (EM) in units $\mathrm{cm^{-5}}$ at those temperature ranges. \textit{White contours} indicate the initial, \textit{green} the reduced CH boundaries. Observation from 14 October 2012.}
		\label{F-EMRANGEOct}
	\end{figure}   
	\FloatBarrier
	\bibliographystyle{spr-mp-sola}
	\bibliography{bibliography}  
\IfFileExists{\jobname.bbl}{} {\typeout{}
\typeout{****************************************************}
\typeout{****************************************************}
\typeout{** Please run "bibtex \jobname" to obtain} \typeout{**
the bibliography and then re-run LaTeX} \typeout{** twice to fix
the references !}
\typeout{****************************************************}
\typeout{****************************************************}
\typeout{}}

\end{article} 
\end{document}